\documentstyle[11pt,emulateapj,psfig]{article}

\def\fnsiz{\footnotesize}

\newcommand{\etal}{{\it et~al.\/}}
\newcommand{\HI}{\mbox {H\thinspace{\fnsiz I}}}
\newcommand{\HII}{\mbox {H\thinspace{\fnsiz II}}}
\newcommand{\Hline}[1]{\mbox{H{\fnsiz {#1}}}}
\newcommand{\Halpha}{\Hline{\mbox{$\alpha$}}}
\newcommand{\Hbeta}{\Hline{\mbox{$\beta$}}}

\begin{document}
\submitted{ApJ, accepted 1999 March 25}
\title{The Taxonomy of Blue Amorphous Galaxies: II. Structure and Evolution}
\author{Amanda T. Marlowe, Gerhardt R. Meurer, \and\ Timothy M. Heckman}
\affil{Department of Physics \& Astronomy, The Johns Hopkins University,
Baltimore, MD 21218 \\ e-mail: 
marlowe@pha.jhu.edu, meurer@pha.jhu.edu, heckman@pha.jhu.edu}
\begin{abstract}

Dwarf galaxies play an important role in our understanding of galaxy
formation and evolution, and starbursts are believed to strongly affect
the structure and evolution of dwarf galaxies. We have therefore
embarked on a systematic study of 12 of the nearest dwarf galaxies
thought to be undergoing bursts of star-formation. These were selected
primarily by their morphological type (blue ``amorphous'' galaxies). We
show that these blue amorphous galaxies are not physically
distinguishable from dwarfs selected as starbursting by other methods,
such as blue compact dwarfs (BCDs) and \HII\ galaxies. All these classes
exhibit surface brightness profiles that are exponential in the outer
regions ($r \gtrsim 1.5r_e$) but often have a predominantly central blue
excess, suggesting a young burst in an older, redder galaxy. Typically,
the starbursting ``cores'' are young ($\sim 10^7-10^8$ yr) events
compared to the older ($\sim 10^9-10^{10}$ yr) underlying galaxy (the
``envelope''). The ratio of the core-to-envelope in blue light ranges
from essentially zero to about two. These starbursts are therefore
modest events involving only a few percent of the stellar mass.

The envelopes have surface-brightnesses that are much higher than
typical dwarf Irregular (dI) galaxies, so it is unlikely that there is a
straightforward evolutionary relation between typical dIs and dwarf
starburst galaxies. Instead we suggest that amorphous galaxies may
repeatedly cycle through starburst and quiescent phases, corresponding
to the galaxies with strong and weak/absent cores respectively. Once
amorphous galaxies use up the available gas (either through
star-formation or galactic winds) so that star-formation is shut off,
the faded remnants would strongly resemble dwarf elliptical
galaxies. However, in the current cosmological epoch this is evidently a
slow process that is the aftermath of a series of many weak, recurring
bursts. Present-day dEs must have experienced more rapid and intense
evolution than this in the distant past.
\end{abstract}

\section{Introduction}

Starbursts, especially in dwarf galaxies, play a key role in
understanding galaxy formation and evolution.  A strong burst of star
formation is expected to occur during the initial virialization and
baryonic dissipational collapse of a proto-galaxy.  In hierarchical
cosmological scenarios, the first systems to virialize (at $z \approx
3$) have masses similar to those of dwarf galaxies, which then
merge to form larger galaxies (e.g.\ \cite{sz78}).  Late epoch
starbursts have been invoked to explain the excess of faint blue
galaxies in the field (\cite{br92}; \cite{bf96}), and the increasing
contribution of blue galaxies with $z$ in clusters (\cite{cs87}).

Starburst driven galactic winds (\cite{ham90}) have a particularly
important role in dwarf galaxy evolution.  Mass loss is easier in the
shallow potential of a dwarf galaxy, so winds may play an important role
in regulating the gas content of dwarfs (\cite{ds86}).  While the
initial geometry of the ISM is also important (\cite{dYH94}), and
extended dark matter halos complicate the mass-loss predictions, it is
unlikely that a dwarf galaxy can retain the hot ($\sim 10^7$ K) metal
enriched gas that inflates the Kpc scale \Halpha\ bubbles commonly seen
in starburst dwarfs (e.g.\ \cite{hdlfgw95}).  Hence, starburst driven
winds, especially from dwarfs, play an important role in enriching the
intergalactic (and intra-cluster) medium, while keeping the metal
content of the remaining ISM lower than closed-box model predictions.
Dwarf galaxies may contribute strongly to QSO absorption line
spectra, especially because of their winds, as well as their
often-extended gaseous disks (e.g.\ DDO154: \cite{cb89}; NGC~2915:
\cite{mcbf96}).

While these issues illustrate the conceptual importance of starbursts
and dwarf galaxies, our understanding of these objects are far from
complete.  Two questions in particular, still are not resolved.  

Firstly, what is the role of starbursts in dwarf galaxy evolution?  In
some models a starburst is a singular event, which through rapid star
formation and consequent galactic wind, transforms a gas rich dwarf
irregular into a low surface brightness, gas poor system (e.g.\
\cite{br92}).  Hence starbursting dwarfs may be the link between gas
rich and gas poor configurations.  However observations of the stellar
populations in even the lowest mass dwarfs show that they typically have
had multi-episodic or an otherwise complex star formation history (e.g.\
\cite{dacosta97}, and references therein).  Given the usual disk
geometry of the ISM, even in dwarfs, and the geometric arguments of De
Young \&\ Heckman (1994\markcite{dYH94}; cf. \cite{mf99}), it
is not clear that winds alone can sweep the ISM out of a dwarf galaxy.
This has also been discussed from an observational point of view by 
Welch {\it et al} (1998).

Secondly, how do starbursts themselves evolve?  The theoretical concept
is that starbursts have a very short duration.  For example the fading
dwarf model of Babul and Ferguson (1996) requires a burst duration $dt
\lesssim 10$ Myr.  However observations of well known starbursts show
them to be complex, consisting of multiple star clusters embedded in a
dominant distribution of diffusely distributed high mass stars
(\cite{mhlkrg95}).  Can such structures be formed on such short
timescales?

In order to address these questions, we have commenced an observational
study of nearby star-bursting dwarf galaxies.  Not only may these
galaxies be the active link between gas-rich and gas-poor dwarf galaxy
configurations but they are the most numerous type of starbursting
systems, hence allowing investigation of relatively nearby
systems (e.g. within the Virgo Cluster distance: D $\leq$ 16 Mpc).

Starbursting dwarf galaxies go by several names.  These include blue
compact dwarf (BCD), \HII\ or emission-line galaxies, and amorphous
galaxies.  These names indicate slightly different selection criteria.
For example Thuan \&\ Martin (1981\markcite{tm81}) define BCDs by low
luminosity, small sizes, blue colors and an emission line spectrum, and
thus is both a color and spectroscopic classification. The \HII\ galaxy
classification is purely spectroscopic, based usually on prism surveys
for emission line sources (e.g.\ \cite{ml78}).  The ``amorphous''
classification introduced by Sandage \&\ Brucato (1979\markcite{sb79}) is
morphological.  It is based on deep photographic imaging and requires an
extended high surface brightness region in a host containing no spiral
arms. Hence amorphous galaxies resemble E or S0 galaxies, but typically
have blue colors.

Our sample of twelve starburst dwarfs ($M_B \gtrsim - 18$) was selected
primarily from blue galaxies classified as amorphous, or likely
amorphous galaxies (by \cite{sb79}; \cite{gh87}; \cite{hvwg94}) because
these galaxies include some of the nearest starburst dwarfs known.  Two
well known dwarf \HII\ galaxies, Haro~14, and II~Zw~40, were also included
to round out our observing time.  The basic properties of the sample
galaxies are listed in Table \ref{tabStats}.  Their proximity makes them
an ideal class for studying starburst galaxies in detail.

\begin{deluxetable}{lccccrcrrrcrrr}
\footnotesize
\tablewidth{0pt}
\tablecaption{Basic Properties of Galaxies\label{tabStats}}
\tablehead{
\colhead{Galaxy}        & \colhead{$v_{\rm helio}$}             &
\colhead{$v_0$}         & \colhead{$D$}                         &
\colhead{scale}         &
\colhead{$r_{e,B}$}     & \colhead{$r_{25}$}                    &
\colhead{$b/a$}         & \colhead{P.A.}                        &
\colhead{$M_{B_0}$}     & \colhead{$\mu_{e,B}$}                 &
\colhead{$A_B$}         & \colhead{[O/H]$_\odot$} \\
\colhead{}              & \colhead{\scriptsize(km s$^{-1}$)}   &
\colhead{\scriptsize(km s$^{-1}$)} & \colhead{\scriptsize(Mpc)}  &
\colhead{\scriptsize(pc $\prime\prime^{-1}$)}      &
\colhead{\scriptsize(pc)} & \colhead{\scriptsize(kpc)}          &
\colhead{}              & \colhead{\scriptsize(deg)}            &
\colhead{}              & \colhead{\scriptsize(mag$/\sq^{\arcsec}$)}  &
\colhead{}              & \colhead{}            \\
\colhead{(1)}           & \colhead{(2)}                         &
\colhead{(3)}           & \colhead{(4)}                         &
\colhead{(5)}           &
\colhead{(6)}           & \colhead{(7)}                         &
\colhead{(8)}           & \colhead{(9)}                         &
\colhead{(10)}          & \colhead{(11)}                        &
\colhead{(12)}          & \colhead{(13)}
}
\startdata
 Haro 14  &\phn944  & 1040 & 12.5 & \phn60 & \phn580 & 2.0 & 0.88 & 50 & --17.02 & 19.65 & 0.24 &\nodata & \nl
 NGC 625  &\phn386 &\phn333 &\phn4.1 &\phn20 &\phn580 & 1.8 & 0.39 & 92 & --16.31 & 21.20 & 0.18 &\nodata & \nl
 NGC 1510 &\phn968 &\phn830 & 11.0 &\phn53 &\phn370 & 1.9 & 0.85 & 90 & --16.75 & 18.93 & 0.16 & --0.7 & \nl
 NGC 1705 &\phn629 &\phn433 &\phn6.1 & \phn30 &\phn260 & 1.4 & 0.75 &  50 & --16.20 & 18.92 & 0.18 & --0.35 & \nl
 NGC 1800 &\phn803 &\phn664 &\phn9.2 & \phn45 &\phn520 & 1.9 & 0.63 & 113 & --16.72 & 19.73 & 0.13 & --0.46 & \nl
 NGC 2101 & 1192 &\phn973 & 13.6 &  66  & 1600 & 2.6 & 0.64 & 85 & --17.40 & 22.24 & 0.45 &\nodata & \nl
 II Zw 40 & \phn789 &\phn751 & 11.1 & \phn54 &  \phn470 & 2.4 & \nodata  & \nodata & --16.77 & 19.42 & 1.86 & --0.8 & \nl
 NGC 2915 & \phn460 &\phn183 &\phn2.9 &\phn14 & \phn190 & 0.7 & 0.54 & 129 & --14.73 & 20.02 & 0.79 & $\le-0.30$ & \nl
 NGC 3125 & 1110 &\phn827 & 13.8 &\phn67& \phn500 & 2.5 & 0.90 & 114 & --18.03 & 18.24 & 0.71 & --0.46 & \nl
 NGC 3955 & 1491 & 1227 & 21.1 & 102 & 1770 & 5.9 & 0.33 & 165 & --19.53 & 21.17 & 0.62 &\nodata & \nl
 NGC 4670 & 1069 & 1031 & 14.6 &\phn70 & \phn 540 & 2.8 & 0.83 &  90 & --17.78 & 18.81 & 0.12 & --0.50 & \nl
 NGC 5253 & \phn404  &\phn156 &\phn3.3 & \phn16 &  \phn350 & 1.8 & 0.50  &  45 & --17.05 & 19.58 & 0.43 & --0.43 & \nl
\enddata
\tablecomments{ (2) Galaxy heliocentric velocity taken from the
NASA/IPAC Extragalactic Database (NED). (3) Velocity relative to the
centroid of the Local Group, computed from the heliocentric velocity in
col.  (2) following the precepts of the RSA. (4) Adopted distance to the
galaxy. For NGC 5253, we have used a distance of 3.3 Mpc for the
Centaurus group. For other galaxies, we used a Virgocentric infall model
(cf. Schecter 1980) with the parameters $\gamma = 2, \omega_\odot = 222$
km s$^{-1}$, $v_{\rm{Virgo}}=976$ km s$^{-1}$ (Bingelli et al. 1986),
and D$_{\rm{Virgo}} = 15.9$ Mpc (i.e., $H_0=75$ km s$^{-1}$
Mpc$^{-1}$). (5) scale in pc per arcsecs. (6) Effective (half-light)
radius of galaxy derived from our B-band images. (7) Isophotal radius at
a surface brightness level of 25 B mag arcsec$^{-2}$ (8) Ratio of
semiminor to semimajor axes at $r_{25}$ dervied from our optical images.
(9) Position angle of the optical major axis at large radii derived from
our optical images. (10) Absolute total B magnitude.  For most cases,
this was determined from the growth curve of the galaxy. NGC 5253 took a
large fraction of the frame, and the growth curve did not level off, so
the flux outside of $r_{25}$ was extrapolated using the exponential fit
and added to the flux inside $r_{25}$ to obtain the total
magnitude. (11) Face on surface brightness within $r_e$ , corrected for
Galactic extinction. (12) Galactic extinction for B (in magnitudes)
based on the HI column densities in Stark et al. 1992, using the
extinction curve provided in Mathis (1990). (13) The log of the
metallicity of the ionized gas relative to solar (Meurer et. al (1994),
Storchi-Bergmann et al. 1995, Telles 1996). }
\end{deluxetable}

The data we collected on these galaxies include broad band UBVI imaging,
narrow band \Halpha\ imaging, and long-slit \Halpha\ spectroscopy.  An
initial paper, Marlowe \etal\ (1995\markcite{mhws95}), presents \Halpha\
imaging and spectroscopy of the seven galaxies in our sample with
evidence of outflows.  It shows that the bubbles have short ($\sim 10$
Myr) expansion timescales, and expansion velocities close to, but
probably smaller than, the escape velocity of the galaxies.  The
energetics of the outflows were estimated and shown to be in accord with
models of the young populations in the centers of these galaxies.  In
the first paper of this series (Marlowe \etal\ 1997; hereafter
\cite{pap1}) we presented the entire dataset on all the galaxies in the
sample.  The broad band images were used to construct surface brightness
profiles, which show that these galaxies typically consist of a young
blue core population, which we identify as the starburst, within an older
redder envelope with an exponential radial surface brightness profile,
which we identify as the underlying ``host'' galaxy.  We also compared
the global properties (e.g.\ magnitudes, colors, \Halpha\ luminosities,
\HI\ fluxes, etc\ldots) of our sample with samples of non-bursting gas
rich dwarf irregulars (dI), gas poor dwarf ellipticals (dE), as well as
other samples of dwarf starbursts (i.e.\ BCD and \HII\ galaxy samples).
While there were clear differences (in the obvious sense) in the global
properties of our sample compared to dEs and dIs, the differences with
other dwarf starburst samples were small and could be traced to the
original selection criteria.  This strengthens the claim that the
differences in the dwarf starburst classifications corresponds to little
more than different naming schemes or slight differences in selection
techniques.

Here we consider the structure and stellar population of our sample in
more detail.  In \S\ref{secData} the data are summarized.  In
\S\ref{secStruct} we consider the structure of our sample galaxies and
show that the structure of most of the galaxies in our sample is
qualitatively and quantitatively the same as found in other dwarf
starburst classifications, confirming that these classifications isolate
the same physical phenomenon.  In \S\ref{secHost} the structure and
stellar content of the underlying envelope, or host galaxy, is used to
test possible evolutionary connections between starburst dwarfs, dIs and
dEs. We also fit models to the core and envelope colors to constrain the
star formation history.  Our conclusions are summarized in
\S\ref{secConc}.

\section{Data and Analysis}\label{secData}

The complete dataset for this project consists of UBVI and \Halpha\
Fabry-Perot images and \Halpha\ \'Echelle spectra.  The reduction and
analysis of these data are discussed in detail in \cite{pap1}, and the
reader is referred there for full details.  Here we are primarily
concerned with the broad band imaging data. The UBVI images were fully
reduced, and calibrated as described in \cite{pap1}. 

\begin{figure*}
\centerline{\hbox{\psfig{figure=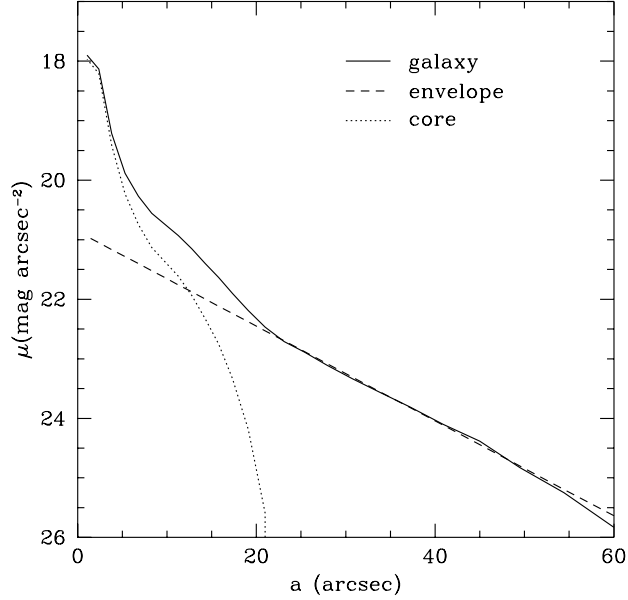,height=8.5cm}}}
\caption{Sample decomposition of NGC 1705's surface brightness
profile into envelope and core components.\label{figSampDecomp} }
\end{figure*}

The surface brightness profiles of most of our sample (8/12) appear to
have two components, an outer region that is roughly exponential, which
we call the envelope, and an inner region in excess of this inwardly
extrapolated exponential, which we call the core. In all cases we fit an
exponential to the outer portion of the galaxies between $1.5 r_{e,B}$
(where $r_{e,B}$ is the effective or half light radius in the $B$ band
of the measured profile), and the outer limits of the photometry (where
the surface brightness is roughly twice the sky level uncertainty).  The
exponential fit has two parameters, the scale length $\alpha^{-1}$, and
the extrapolated face-on central surface brightness $\mu_{0,c}$
(Freeman, 1970).  For a pure exponential profile $r_e = 1.68\alpha^{-1}$
and $\mu_e = \mu_{0,c} + 1.12$ mag.  We subtract the exponential model
from the surface brightness profile, which yields the core surface
brightness profile. Figure \ref{figSampDecomp} shows the decomposition
of NGC 1705's surface brightness into the core and envelope components
as an example.  Comparisons of the envelope and the core can be found in
Table \ref{tabDCProp}. As Haro 14 and NGC 3955 have almost pure
exponential profiles, the results for the core are highly
uncertain. Therefore we present only their envelope properties and the
inner colors (within $0.5r_{e,B}$) of the entire galaxy.  Neither
envelope-core decomposition, nor exponential fit, were done for NGC~2101
due to the poor quality of the outer profiles, resulting from scattered
light from a bright foreground star.  Here we present only the ``core''
properties as defined by the inner colors within $0.5r_{e,B}$.

\begin{deluxetable}{lccrccccccrccc}
\footnotesize
\tablewidth{0pc}
\tablecaption{Envelope vs. Core: Basic properties\label{tabDCProp}}
\tablehead{
&\multicolumn{6}{c}{Envelope} & { } & \multicolumn{4}{c}{Core} \\
\cline{2-7} \cline{9-12} \\
\colhead{Galaxy}&
\colhead{M$_B$} &\colhead{$\mu_{0,c}$} & \colhead{$\alpha^{-1}$}&
\colhead{(U$-$B)}& \colhead{(B$-$V)} & \colhead{(V$-$I)}& &
\colhead{M$_B$} &
\colhead{(U$-$B)}& \colhead{(B$-$V)} & \colhead{(V$-$I)} \\
\colhead{}&
\colhead{} &\colhead{\scriptsize(mag arcsec$^{-2}$)} & \colhead{\scriptsize(pc)} &
\colhead{} &\colhead{} & \colhead{} &&
\colhead{} &
\colhead{} &\colhead{} & \colhead{}\\
\colhead{(1)}&
\colhead{(2)} &\colhead{(3)} & \colhead{(4)} &
\colhead{(5)} &\colhead{(6)} & \colhead{(7)} &&
\colhead{(8)} &\colhead{(9)} & \colhead{(10)} &
\colhead{(11)} 
}
\startdata
Haro 14  & --17.04 & 19.7 &  430 &   0.03 & 0.51 & 0.38 & &\nodata & (--0.29) & (0.31) & (0.54)\nl
NGC 625  & --16.28 & 21.3 &  650 & --0.01 & 0.55 & 0.84 & & --14.71 & --0.49 &   0.22 & 0.49 \nl
NGC 1510 & --16.17 & 20.9 &  530 & --0.04 & 0.51 & 0.78 & & --15.97 & --0.37 &   0.24 & 0.39 \nl
NGC 1705 & --15.57 & 21.0 &  390 & --0.23 & 0.48 & 0.69 & & --15.53 & --0.84 & --0.03 & 0.50  \nl
NGC 1800 & --16.28 & 21.3 &  610 & --0.05 & 0.56 & 0.74 & & --15.77 & --0.32 &   0.15 & 0.43 \nl
NGC 2101 &\nodata &\nodata &\nodata &\nodata &\nodata &\nodata &&\nodata & (--0.49) & (0.35) & (0.16) \nl
NGC 2915 & --14.43 & 20.7 &  200 &   0.04 & 0.50 & 0.92 & & --13.36 & --0.59 &   0.23 & 0.53 \nl
NGC 3125 & --17.54 & 19.5 &  480 & --0.34 & 0.35 & 0.63 & & --17.42 & --0.45 &   0.31 & 0.71 \nl
NGC 3955 & --19.38 & 20.9 & 2170 &   0.07 & 0.71 & 0.78 & &\nodata & (--0.17) & (0.45) & (0.93) \nl
NGC 4670 & --17.38 & 20.3 &  690 & --0.27 & 0.45 & 0.76 & & --17.39 & --0.60 &   0.29 &\nodata \nl
NGC 5253 & --16.57 & 20.7 &  550 & --0.07 & 0.46 & 0.76 & & --16.16 & --0.65 &   0.22 & 0.44 \nl
\enddata
\tablecomments{ Col(2)-- Absolute total B magnitude of the envelope,
corrected for Galactic extinction. Col(3)--Face on envelope B band central
surface brightness of the underlying exponential galaxy, corrected for
Galactic extinction, derived from the exponential fit. Col(4) Scale
length in the B band of the envelope derived from the exponential fit.  Cols
(5),(6),(7)--Total exponential envelope colors, corrected for Galactic
extinction.  Col(8)--Absolute total B magnitude of the ``core'' (the
residual galaxy after the exponential envelope described above is
subtracted), corrected for Galactic extinction.  Cols
(9),(10),(11)--Total core colors, corrected for Galactic extinction and
emission-line contamination.  Colors shown in parentheses are not based
on a core-envelope decompostion, but are merely the inner colors ($r <$
0.5$r_{e,B}$)}.
\end{deluxetable}

We caution that the core-envelope separation is prone to large
systematic errors.  Consider first the effect of the apertures used to
extract the profiles.  In Paper I (appendix) we showed that fits to
surface brightness profiles derived from circular annuli when applied to
a galaxy with elliptical isophotes will underestimate the central
surface brightness and overestimate the scale length. Such an error will
also effects the core/envelope ratio. In the worst case, Haro 14 appears
to have a core (core/envelope $\sim 0.4$) in the circular annuli
profile, but little or no core when elliptical annuli photometry is used
(Paper I, fig.~7). We do considerably better by using elliptical annuli
photometry, because the isophotes are better approximated by ellipses.
Using variable ellipse parameters fitted to the isophotes yields
profiles of higher fidelity (with respect to the isophotes) than those
from averaged parameters. However it results in a conceptual problem: if
the shape varies with radius what is the shape of the envelope where the
core dominates?  For our sample the difference in total flux of the
envelope is less than 15\%\ when results from variable and constant
shaped ellipse parameters are compared, and there is no bias in the sign
of the difference.  Hence we adopt the constant shaped elliptical annuli
photometry for the exponential fits and the core-envelope
separation. There is one exception to this: II~Zw~40 which is a merging
dwarf system (\cite{bk88}; \cite{vzss98}) and has an ``$\times$''
morphology (\cite{pap1}).  For it we adopt the concentric circular
annuli profiles centered on the brightness peak.  The radial fit range
(set by the depth of the photometry) also effects the results.  As a
worst case consider NGC~2915, for which we derive $F_{\rm B,core}/F_{\rm
B,env} = 0.37$ ($B$ band core/envelope flux ratio) from our photometry.
Using the deeper photometry of Meurer et al.\ (1994\nocite{mmc94}) we
derive $F_{\rm B,core})/F_{\rm B,env} = 0.89$.  In this case the onset
of the core is gradual, and the fit we present in \cite{pap1} includes
part of the core seen in the Meurer et al.\ profile.  Better agreement
is found for NGC~1705, for which we derive $F_{\rm B,core}/F_{\rm B,env}
= 0.96$, whereas Meurer et al.\ (1992) have $F_{\rm B,core}/F_{\rm
B,env} = 0.91$.  In summary, mismatches in isophote shapes could cause a
``random'' error of $\lesssim 0.15$ in $F_{\rm B,core}/F_{\rm B,env}$,
while the limited depth of our photometry may result in the ratio being
underestimated by $\lesssim 0.5$ in some cases.  The latter effect will
be smaller in the $I$ band where the core is weaker.

As we are using the galaxy colors to make quantitative statements about
the galaxy's age, we need to consider the effects of emission line
contamination. While in some galaxies the corrections are negligible, in
others, colors can be affected by as much as a few tenths of a
magnitude.  UV-optical spectra for six of our twelve galaxies, kindly
provided by Daniela Calzetti (see \cite{calzetal}), were used to measure
the equivalent widths of the emission lines that fall within each of our
filters for these galaxies. The aperture size of the spectra is
$10^{\prime\prime} \times 20^{\prime\prime}$, which encompasses most of
the cores in the galaxies. Thus the correction would only be valid in
the inner regions of the galaxy, so we only apply them to the
core. Excluding the exceptional case of NGC 1705, the average value for
the ratio of the half-light radii of the emission-line gas and optical
continuum in these galaxies is about 50\% (see Paper I for details).
Since the gas is usually more compact than the starlight, we therefore
make no corrections to the envelope colors.

Comparing the equivalent widths with
the FWHM of the filter, and adjusting for throughput differences of
the different emission lines, we estimate a correction for each
filter:

\begin{equation}
\label{eqEMLCor}
\Delta m = 2.5 \log\left( 1 + \left(\sum_i{\frac{EW_i}{{\rm
FWHM}_{fil}}
\frac{T_i}{T_{fil}}} \right)\right),
\end{equation}
where $\Delta m$ is the correction to filter $m$ in
magnitudes, $EW_i$ is the equivalent width of emission line $i$ found
in filter $m$, FWHM$_{fil}$ is the full width at half maximum of filter
$m$, ${T_i}$ is the throughput of the filter at the wavelength of
emission
line $i$, and $T_{fil}$ is the average throughput of the filter within
its FWHM bandpass.

If we assume the corrections scale with the \Halpha\ equivalent width
(derived from our \Halpha\ images), and that NGC 5253 and NGC 3125 have
line ratios that are typical of the sample, we can estimate the
corrections for those galaxies for which we do not have spectra. Of the
remaining galaxies, NGC~2101, Haro~14, and II~Zw~40 have \Halpha\ 
equivalent widths that suggest emission line contamination would be a
significant problem. For Haro~14 and NGC~2101 we estimate the
corrections by assuming they have spectra similar to NGC 3125. II~Zw~40,
however, has an exceptionally high \Halpha\  equivalent width ($\sim
400$\AA\ in the central region), and may have very different line
ratios. Therefore we do not have much confidence in the accuracy of such
corrections, and feel it best to make no attempt at correcting II Zw 40
without multiwavelength spectra. The final corrected core colors for
Haro 14, NGC 2101, NGC 3125 and NGC 5253 are given in Table
\ref{tabDCProp}. The Leitherer \&\ Heckman (1995) stellar population
synthesis models used below include the contribution from the nebular
continuum (see Leitherer \&\ Heckman 1995\markcite{lh95} text for
discussion); therefore we do not correct for nebular continuum
contamination.

\section{The Structure of Blue Amorphous Galaxies}\label{secStruct}

\subsection{Morphology and Structure}\label{ssecMorph}

Our sample, primarily selected to consist of nearby blue dwarf amorphous
galaxies, includes galaxies with a variety of morphologies.  The
dominant form is galaxies with smooth round outer isophotes and
centrally concentrated star formation.  These are iE or nE galaxies in
the nomenclature Loose \&\ Thuan (1985\markcite{lt85}) adopt for
describing BCDs.  Indeed they find that for BCDs these are the most
common types.  Our sample also includes II~Zw~40 a BCD/\HII\ galaxy with
irregular structure at all radii (type iI following \cite{lt85}).  Also
in our sample are two highly inclined disk galaxies (NGC~625, NGC~3955)
and one galaxy (NGC~2101) with a morphology like a normal dwarf
irregular, albeit of somewhat high surface brightness.

As we will discuss below, (U--B) colors bluer than about --0.4 are
indicative of a strong starburst (burst age younger than $\sim$ 10$^8$
years for continuous star-formation and younger than $\sim$ 10$^7$ years
for an instantaneous burst). We find that this signature of a starburst
is in most cases correlated with the galaxy surface brightness profile:
galaxies whose core color (U--B) color indicates a starburst have the
afore-mentioned core-envelope structure. Those that do are NGC~1705,
NGC~2915, NGC~3125, NGC~4670, and NGC~5253. Moreover, in all these cases
the core is bluer than the envelope, as can be seen in
table~\ref{tabDCProp}, and the color profiles shown in \cite{pap1}
(Fig.~9).  Four of these galaxies have been {\em firmly\/} classified as
amorphous by Sandage \&\ Brucato (1979\markcite{sb79}) or Gallagher \&\
Hunter (1987\markcite{gh87}).  The exception is NGC~4670 which was
classified amorphous in a later paper (\cite{hvwg94}). These galaxies
all have high effective surface brigthnesses as well: Table 1 
shows that $\mu_{e,B}$ is brighter than 20 mag arcsec$^{-2}$ in all
cases.

Of the remaining five galaxies with redder cores in (U--B), one - Haro
14 - has an almost purely exponential profile; one - NGC~2101 - has a
central plateau (flattening) of its surface brightness profile; one -
NGC~625 - has both a core-halo structure and a central plateau, and two
- NGC~1510 and NGC~1800 have simple core-envelope structures.  Of these,
only NGC~1510 and NGC~1800 are amorphous galaxies (the others are all
border-line amorphous galaxies judging by doubts expressed by Sandage
\&\ Brucato [1979] or Gallagher \&\ Hunter [1987]). NGC~625 and NGC~2101
have considerably fainter effective surface brigthnesses than the other
galaxies ($\mu_{e,B}$ = 21.2 and 22.2 mag arcsec$^{-2}$ respectively -
see Paper I).

Note that we exclude from the above discussion NGC~3955, which is much
more luminous than the other galaxies ($M_B = -19.5$), shows spiral arms
(contrary to the original definition of the Amorphous classification
(\cite{sb79}), and whose central region is riddled by dust lanes whose
associated obscuration affect its colors and radial surface brightness
profile (inducing an apparent plateau).

We conclude that photometric indicators of a starburst (the presence of
a photometrically-distinct core whose colors are bluer than the envelope
and are sufficiently blue to require a short-duration starburst) are usually
in satisfactory agreement with one another. These characteristics are
present with a high incidence rate among bona fide amorphous galaxies.

\subsection{Connection with Outflows}

In Marlowe {\it et al} (1995) we discussed H$\alpha$ \'echelle
spectroscopy that provided kinematic evidence for outflowing gas in
seven of the galaxies in our present sample (NGC~1705, NGC~1800,
NGC~2915, NGC~3125, NGC~3955, NGC~4670, and NGC~5253). In contrast, the
galaxies Haro~14, NGC~625, NGC~1510, NGC~2101, and II~Zw~40 did not
reveal outflows.

Is there some connection between the presence or absence of an outflow
and the structure or stellar content of the galaxy (based upon our
multi-color surface-photometry)? We exclude from consideration the
luminous spiral NGC~3955 (see above) and II~Zw~40 (for which we do not
have reliable colors). Of the remaining ten galaxies, the five with high
effective surface brightnesses ($\mu_{e,B} \leq$ 20 mag arcsec$^{-2}$)
and whose cores have ``starburst colors''((U--B) $< -0.4$) are all
driving outflows of ionized gas (NGC~1705, NGC~2915, NGC~3125, NGC~4670,
and NGC~5253). In marked contrast, only {\it one} of the five galaxies
with lower effective surface brightnesses and/or redder cores is driving
a weak outflow (NGC~1800).

This suggests that the condition needed for driving a large-scale outflow of
ionized gas in a gas-rich dwarf galaxy is
a transient (10 to 100 Myr) episode of star-formation
in the galaxy core in which the rate of star-formation per unit area
is also high. Both the short timescale and high intensity of the
star-formation are in accord with the usual definitions of starbursts
and with theoretical understanding of starburst-driven outflows
(cf. MacLow \& Ferrara 1998).

\subsection{The Dwarf Starburst Galaxy Connection}\label{ssecEvolve}

Here we compare our sample to other samples of dwarf starburst galaxies,
in order to determine to what degree our sample is representative of the
dwarf starburst phenomenon.  The comparison will primarily be with the
samples of Papaderos \etal\ (1996a,b; hereafter P96) and Telles
(hereafter T97: \cite{tmt97}; \cite{tt97}).  The
P96 sample was primarily selected from the Thuan \&\ Martin (1981) list
of BCDs, while the T97 sample was selected from {\em The
Spectrophotometric Catalog of \HII\ Galaxies\/} (Terlevich, 1991).  Hence
we are comparing our morphologically selected sample to samples of color
and emission line selected dwarf starbursts.  

\begin{figure*}
\centerline{\hbox{\psfig{figure=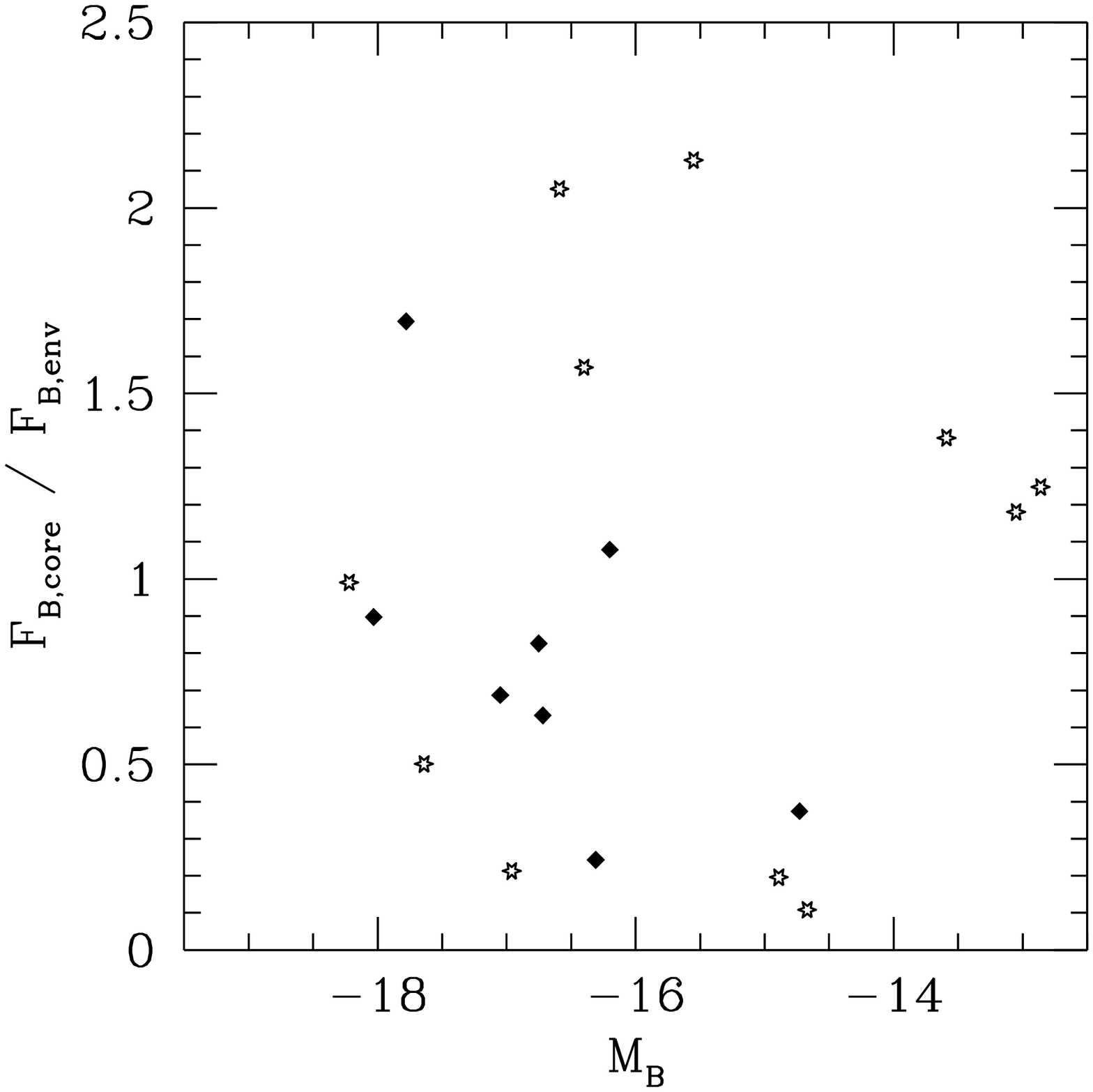,height=8.5cm}
\psfig{figure=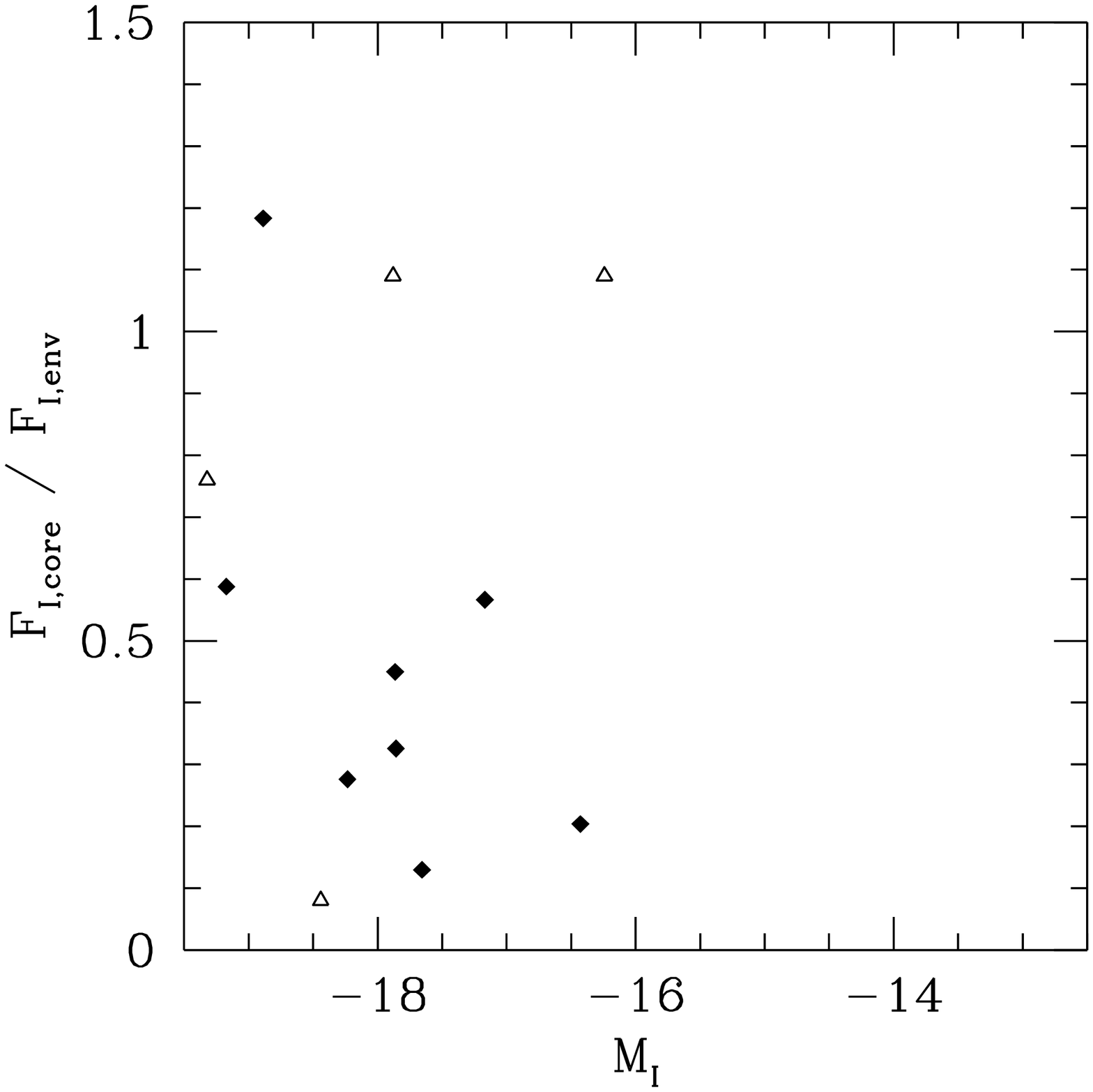,height=8.5cm}}}
\caption{Core to envelope flux ratio plotted against absolute
magnitude for starburst dwarfs. a) $B$ band values: blue amorphous
galaxies (BAGs, diamonds) and BCDs (stars). b) $I$ band values: BAGs
(diamonds) and HII galaxies (triangles).  The uncertainty in these
ratios is likely to be $\lesssim 0.15$ due to mismatches in isophote
shape, while the limited depth of the photometry may result in some
ratios being underestimated by $\lesssim 0.5$, in the $B$ band, with a
lesser effect in the $I$ band.
\label{figCoreFluxRatio}}
\end{figure*}

One concern is that the morphological selection may result in the less
extreme, or older, starbursts since emission lines are not as important
in the selection.  If so, we may expect to see a difference in the burst
strength, as parameterized by the core/envelope flux ratio, then seen in
our sample.  We may also expect our sample to have redder core
colors. Figure \ref{figCoreFluxRatio} compares the ratio of the core to
envelope fluxes of our sample in B versus $M_B$ against the P96 sample,
and in I versus $M_I$ against the T97 sample.  For the T97 sample the I
core flux is estimated by subtracting the total luminosity derived from
the exponential fit to the envelope from the galaxy total luminosity in
those galaxies that showed a core component.  While the I band is less
sensitive to the massive stellar population, it is the only band in
common with our sample for which the profile fits are available in the
T97 sample.  Note the surface brightness profiles of the T97 sample were
extracted using circular annuli and this may artificially induce an
apparent central excess (`core') in the surface-brightness profile (see
\cite{pap1} Appendix for details).

Figure~\ref{figCoreFluxRatio} shows two things.  Firstly, while some
galaxies in the P96 and T97 samples have a stronger core/envelope flux
ratio, there is nevertheless a strong degree of overlap between the
samples.  Secondly, the cores do not totally dominate the galaxy.  The
cores in our sample galaxies with core-envelope structures amount to 0.2
to 1.1 mag brightening in the B band light, relative to the envelope
alone.  In the P96 sample the maximum ratio, 2.15, corresponds to a 1.25
mag enhancement relative to the envelope alone.  Similarly, Sudarsky \&\
Salzer (1995\markcite{ss95}) report core enhancements of 0.2 - 0.6 mag
in their sample of BCDs. Hence starbursts in dwarf galaxies represent
only a modest $\lesssim$ 1 magnitude enhancement to their pre-burst
progenitors.  Even allowing for some downwards biasing in our
core/envelope ratios due to the limited depth of our photometry (see
\S\ref{secData}) the cores represent a $\lesssim$ 1.5 magnitude
enhancement. 

\begin{figure*}
\centerline{\hbox{\psfig{figure=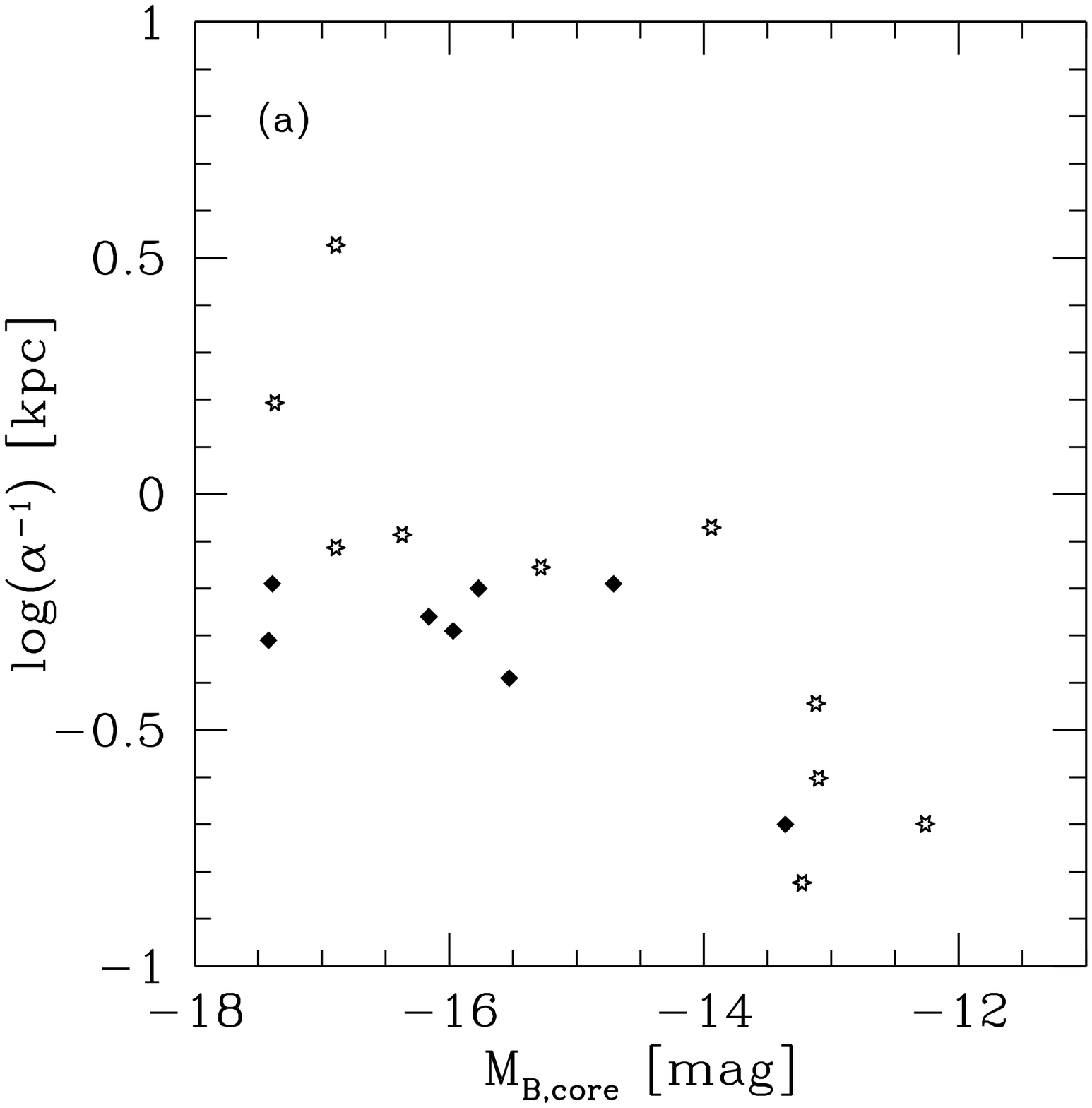,height=8.5cm}
\psfig{figure=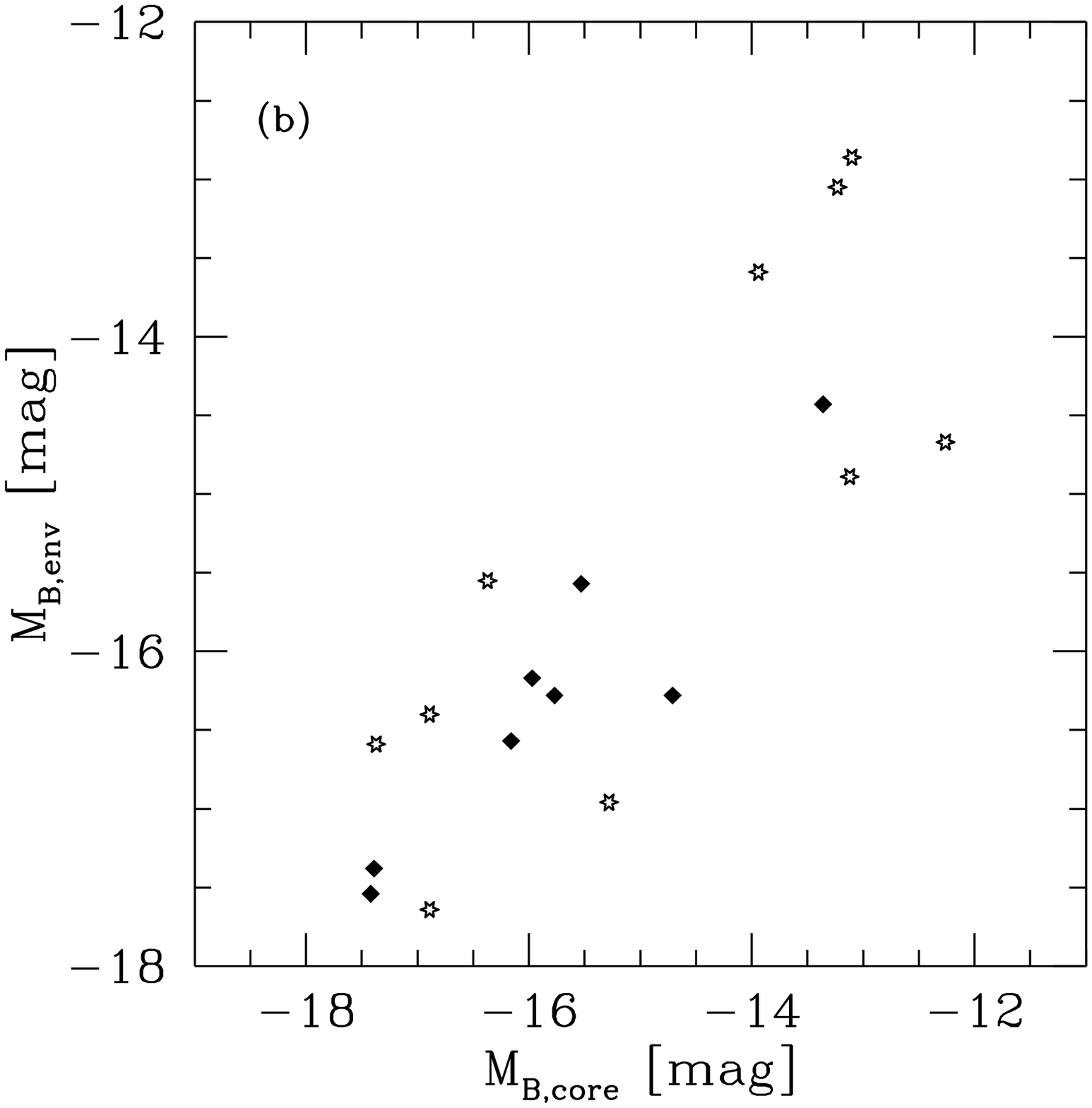,height=8.5cm}}}
\caption{Comparison of cores and envelopes.  a) Envelope scale length
plotted against $M_B$ of the core.  b) $M_B$ of Envelope plotted against
$M_B$ of the core.
\label{figCoreEnv}}
\end{figure*}

In Figure \ref{figCoreEnv}a we compare the absolute magnitude of the
core with the size (scale length) of the envelope for our sample and the
P96 sample. Here the two samples form a sequence, with brighter cores
forming in larger galaxies.  Similarly, Figure \ref{figCoreEnv}b shows
that brighter cores are associated with brighter underlying galaxies
(envelopes). To some degree, both these correlations reflect a selection
effect: bright or large underlying galaxies with weak or absent
starburst cores certainly exist (even in our sample - e.g.  NGC~625,
Haro~14, NGC~2101), but are harder to distinguish. Thus, the absence of
points in the upper right (lower right) of Figure \ref{figCoreEnv}a
(Figure \ref{figCoreEnv}b) is a selection effect. On the other hand, the
absence of very bright cores in small or faint galaxies is not a
selection effect. Thus, these figures suggest that the maximum strength
of the starburst is correlated with the size and luminosity of the host
galaxy. Similar trends are present in samples of more powerful
starbursts in brighter and more massive host galaxies (cf. Lehnert \&
Heckman 1996; Heckman et al 1998).

\begin{figure*}
\centerline{\hbox{\psfig{figure=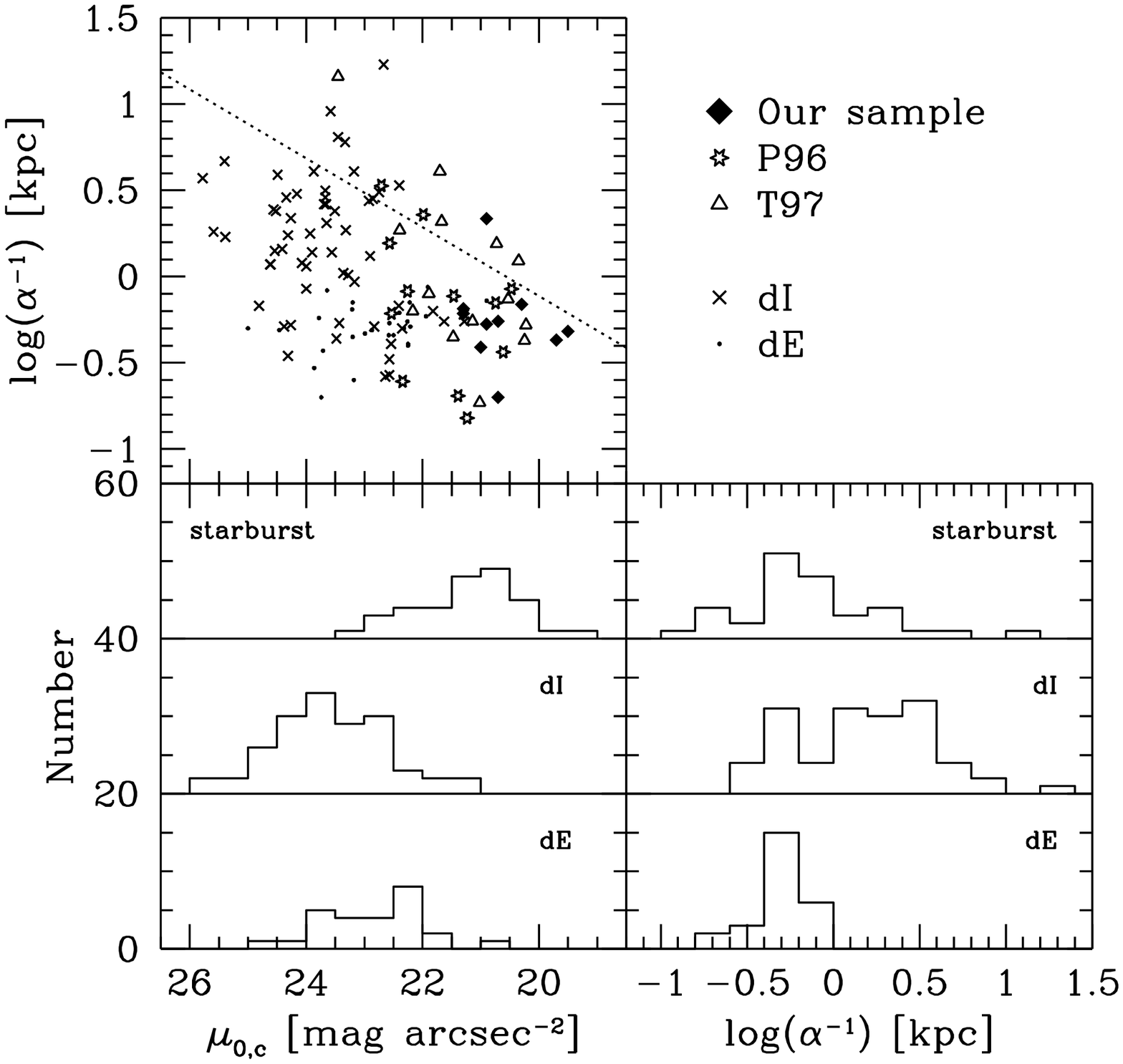,height=8.5cm}}}
\caption{Structural properties of the exponential portion of surface
brightness profiles of (mostly) dwarf galaxies, are shown in the upper
left panel.  The upper right panel shows the symbol correspondence, with
the top three entries corresponding to the {\em envelopes} of starburst
dwarfs: our sample, \protect\cite{pltf96}, and T97; also shown are dIs
-- Patterson \&\ Thuan 1996 (PT)\protect\markcite{pt96}. dEs --
\protect\cite{cb87}.  The quantities are measured in the B band except
for those from T97 which are measured in the I band (see the text for
our adopted transformation).  The dotted line marks the parameters of
exponential profiles having $M_B = -18$ mag.  Galaxies above this line
are too bright to be dwarfs.  The lower two panels project the points
into one dimensional distributions.
\label{figDiskProps}}
\end{figure*}

\begin{figure*}
\centerline{\hbox{\psfig{figure=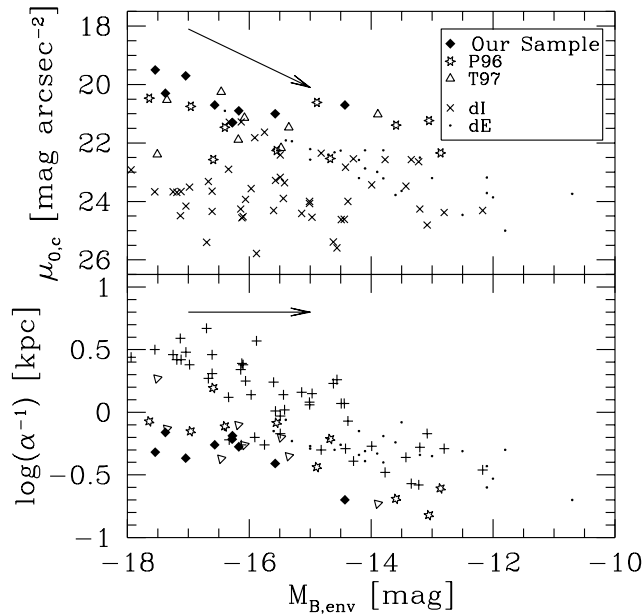,height=8.5cm}}}
\caption{Structural properties of the exponential portion of surface
brightness profiles of dwarf galaxies as a function of absolute blue
magnitude. Samples and symbols as in Fig. \protect\ref{figDiskProps}.
The quantities are measured in the B band except for those from T97
which are measured in the I band (see the text for our adopted
transformation).  The arrows indicate the fading of the envelope if star
formation were cut-off for 10 Gyr (see text). The starbursts could fade
into dEs, but not into the dIs, which are too blue for such fading.
\label{figDiskProps2}}
\end{figure*}

Comparing core colors, we see that the cores in our sample have a mean
$\langle {\rm V-I}\rangle = 0.39$ (Table \ref{tabDCProp}) similar to
that found in the sample of T97 $\langle {\rm V-I}\rangle = 0.45$.  The
envelopes of our sample have similar colors ($\langle {\rm V-I}\rangle =
0.73$) to the envelopes studied by T97: $\langle {\rm V-I}\rangle =
0.76$.  Figure \ref{figDiskProps} compares the structural parameters of
the exponential envelope ($\alpha^{-1}$, $\mu_{0,c}$) of various
starburst dwarf samples as well as other samples of dwarf systems.  In
Figure \ref{figDiskProps2}, we show these parameters as a function of
the absolute magnitude of the exponential envelope. For the T97 sample,
exponential fits were only done in the I band. Thus for these objects we
assume the B scale length is the same as the I scale length, and convert
I magnitude values to B by assuming an average envelope color of
(B--I)$=1.4$. Examination of our sample implies that these assumptions
are reasonable ones For our own sample, we find that the B and I scale
lengths agree well (the B scale lengths are systematically smaller, but
only by an average of 9\%). The (B--I) colors of the envelopes range
from 0.9 to 1.5 with a median of 1.3. On the scale of Figure
\ref{figDiskProps} and Fig.~\ref{figDiskProps2}, these possible offsets
between B and I values are unimportant. A more significant problem with
the T97 sample is that the structural parameters for this sample were
derived from circular aperture profiles.  This biases the results
yielding fainter central surface brightnesses and longer scale lengths
relative to fits of surface brightness profiles extracted through
matched elliptical annuli.  The magnitude of the effect is likely to be
less than 40\%\ in scale length, but up to 1.2 mag in $\mu_{0,c}$
(\cite{pap1} Appendix).

Figure \ref{figDiskProps} and Fig.~\ref{figDiskProps2} show that the
different starburst dwarf samples cover an overlapping area on this
diagram, that is distinct from dI and dE galaxies (more on this in
\S\ref{secHost}).  Our sample tends to have central surface brightnesses
somewhat brighter (by about 0.5 mag arcsec$^{-2}$ in the mean) compared
to the other starburst dwarf samples.  This may in large part be due to
differences in measurement techniques, and may also reflect the greater
mean distances of the T97 galaxies compared to ours (differences in the
mean distances of the various subsamples may result in a systematic
change in the measured parameters since the more distant galaxies are
less well resolved).  As mentioned the T97 sample may have
systematically fainter $\mu_{0,c}$ due to their use of circularly
averaged surface brightness profiles.  In addition our images have
somewhat limited depth and field of view.  Hence the exponential fits
are more weighted to small radii.  If the edge of the core contaminates
the fit there will be a bias towards high $\mu_{0,B}$ (\S\ref{secData}).

\begin{figure*}
\centerline{\hbox{\psfig{figure=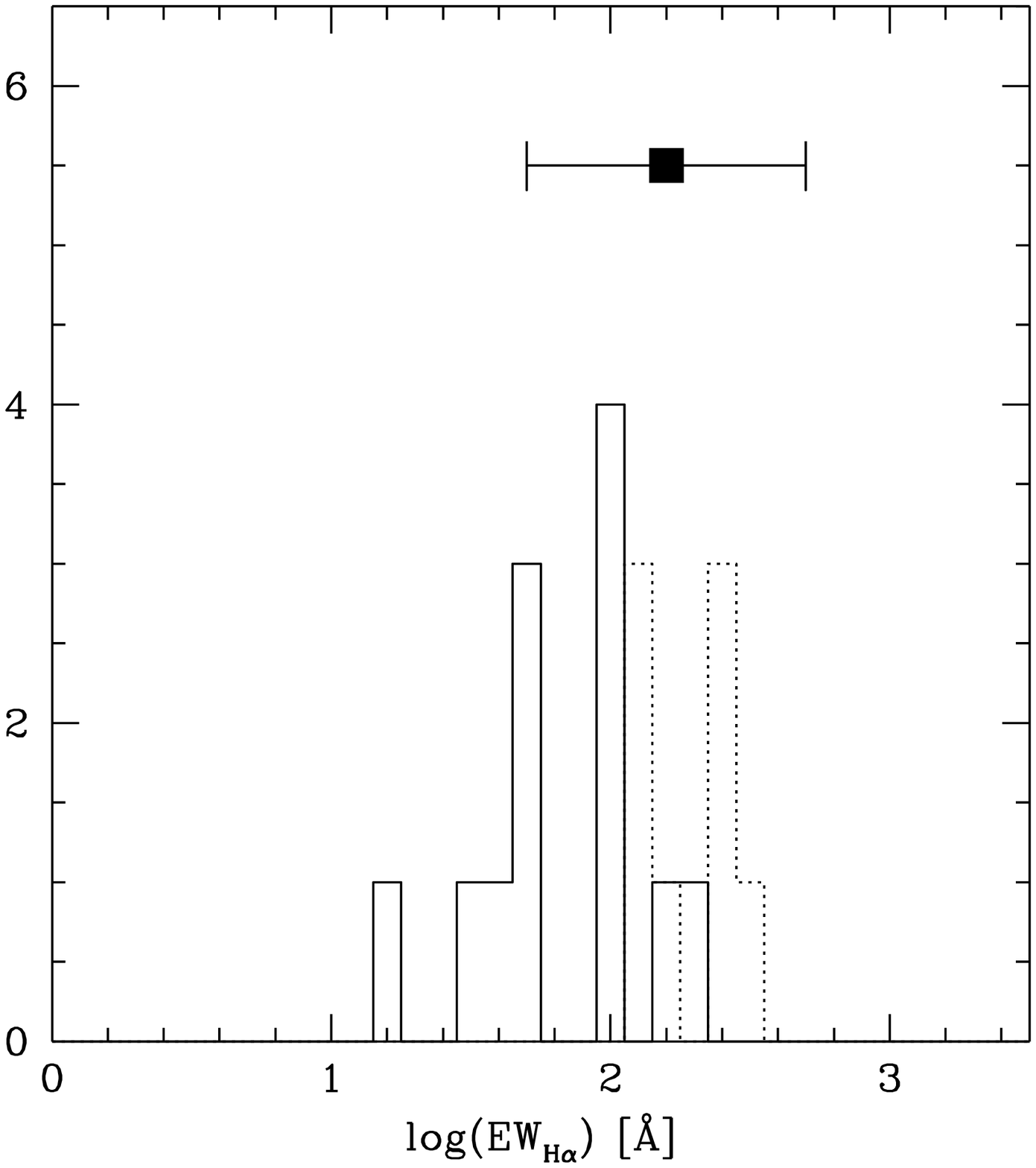,height=8.5cm}}}
\caption{Comparison of H$\alpha$ equivalent widths.  The solid histogram
shows the distribution for the total galaxy H$\alpha$ equivalent width,
while the dotted histogram shows the H$\alpha$ equivalent width assuming
that all H$\alpha$ flux is associated with the core (where a core is
present).  Also shown is the average (large square) and standard
deviation (error bars) of the H$\alpha$ equivalent width for HII
galaxies (\protect\cite{terlea91}).
\label{figEWComp}}
\end{figure*}

Figure \ref{figEWComp} compares the \Halpha\ equivalent widths of our
sample to that of \HII\ galaxies to further investigate core age
differences. The histogram shows our sample while the point with error
bars shows the average and standard deviation of \Halpha\ equivalent
widths given in {\em The Spectrophotometric Catalog of \HII\ Galaxies\/}
(\cite{terlea91}). We note that the Terlevich et al. equivalent widths
were mostly taken through apertures corresponding to $\sim 0.5$ kpc,
which would more likely correspond to the core than to the entire
galaxy, so we show our core upper limit \Halpha\ equivalent widths (see
Table \ref{tabEWComp}) in a dashed line for comparison. As can be seen,
our sample and the \HII\ galaxy sample are broadly similar, although the
latter typically have somewhat higher equivalent widths.  Note that the
subsample of \HII\ galaxies studied by T97 are selected to have the
highest \Hbeta\ equivalent widths, and so should have somewhat younger
luminosity-weighted mean ages than the more comprehensive sample of
Terlevich et al. (1991).

\begin{deluxetable}{lrrrrr}
\tablewidth{0pt}
\tablecaption{H$\alpha$ Equivalent Widths: modeled vs. measured
\label{tabEWComp}}
\tablehead{
\colhead{Galaxy} & \colhead{$EW_{\rm LH,B}$} &
\colhead{$EW_{\rm LH,C}$} & \colhead{$EW_{\rm core}$} &
\colhead{$EW_{\rm spec}$}  & \colhead{$EW_{\rm gal}$} \\
\colhead{(1)} & \colhead{(2)} &
\colhead{(3)} & \colhead{(4)} &
\colhead{(5)} & \colhead{(6)}
}
\startdata
 Haro 14   &   9 & 159 & \nodata & \nodata & 109  \nl
 NGC 625   &  40 & 177 &     243 & \nodata &  45  \nl
 NGC 1510  &   9 & 159 &     127 &      70 &  54  \nl
 NGC 1705  & 136 & 604 &     249 &      42 & 101  \nl
 NGC 1800  &   1 & 159 &     128 &      20 &  42  \nl
 NGC 2101  & 184 & 255 & \nodata & \nodata &  45  \nl
 NGC 2915  &  40 & 177 &     136 & \nodata &  35  \nl
 NGC 3125  &  40 & 177 &     301 &     218 & 168  \nl
 NGC 3955  &   9 & 159 & \nodata & \nodata &  14  \nl
 NGC 4670  &  27 & 177 &     149 & \nodata &  96  \nl
 NGC 5253  &  64 & 177 &     234 &     475 &  92  \nl
\enddata
\tablecomments{(2) EW (in \AA)
predicted by the LH95 instantaneous
burst models for each galaxy, based upon the ages in col (6) of 
Table \protect{\ref{tabAges}}. 
(3) EW (in \AA)
predicted by the LH95 continuous star formation models, based upon
the ages in col (9) of Table \protect{\ref{tabAges}}. 
(4) EW of the
core, assuming that the entire H$\alpha$ flux is emitted by the
burst in the core. The continuum level is determined by
interpolating between the V and I fluxes of the core only, and thus
provides an upper limit on the true core equivalent width. (5) EW
as measured for the inner region of the galaxy, from the
multiwavelength spectra provided by D. Calzetti, taken in a
$10^{\prime\prime} \times 20^{\prime\prime}$ aperture. This would
roughly correspond with the core in most cases (NGC 5253 has a
significantly larger core area), and provides a comparison with the
upper limit given in col (4). (6) EW of the galaxy as a whole. This
is determined by dividing the total H$\alpha$ flux by the R flux of
the entire galaxy. Again, the continuum level was found by
interpolating between the V and I fluxes of the galaxy. }
\end{deluxetable}

Thus we conclude that our sample of predominantly morphologically
selected dwarf starburst galaxies, i.e., blue amorphous galaxies,
represent the same basic physical phenomena as other dwarf starburst
samples, i.e.\ BCDs and \HII\ galaxies. There is a substantial
overlap in such key paremeters as core colors,
envelope colors, \Halpha\ equivalent widths, core to envelope flux
ratios, and envelope structural parameters.  We note only one difference
with other dwarf starburst samples: BCDs and \HII\ galaxies tend to have
somewhat lower metallicities (oxygen abundance) than amorphous galaxies
($\sim 1/6Z_\odot$ versus $\sim 1/3Z_\odot$; \cite{pap1}).  This is
probably a second order effect and we do not consider it any further
here.

\subsection{Star Formation Timescales of Cores and 
Envelopes}\label{secCETimes}

We estimate ages for the cores and envelopes by comparing the UBVI
colors of the cores and envelopes with the models of Leitherer \&\
Heckman (1995, hereafter \cite{lh95}) and Bruzual \& Charlot (1996,
hereafter \cite{bcmod93}, see Bruzual \& Charlot 1993 for details).  The
latter models extend to larger timescales ($t > 4 \times 10^8$) years,
which turns out to be important for modeling the envelopes. Note that
the relatively large difference between the \cite{lh95} and
\cite{bcmod93} evolutionary tracks at ages $\leq 10$ Myr is due to the
fact that \cite{lh95} take into account the contribution of the nebular
continuum, whereas \cite{bcmod93} does not. The LH95 models we use
employ a Salpeter IMF ranging from 0.1$M_\odot$ to an upper cutoff mass
of 100 $M_\odot$. We interpolate between the models for solar and
quarter solar metallicities to obtain a model of one third solar
metallicity, the average metallicity of our sample (see Table
\ref{tabStats}). These models are used to determine the core
properties. The \cite{bcmod93} models used here employ a Salpeter IMF
ranging from $0.1M_\odot$ to $100M_\odot$ and have 40\%\ solar
metallicity.  They are employed to determine the envelope properties.

\begin{figure*}
\centerline{\hbox{\psfig{figure=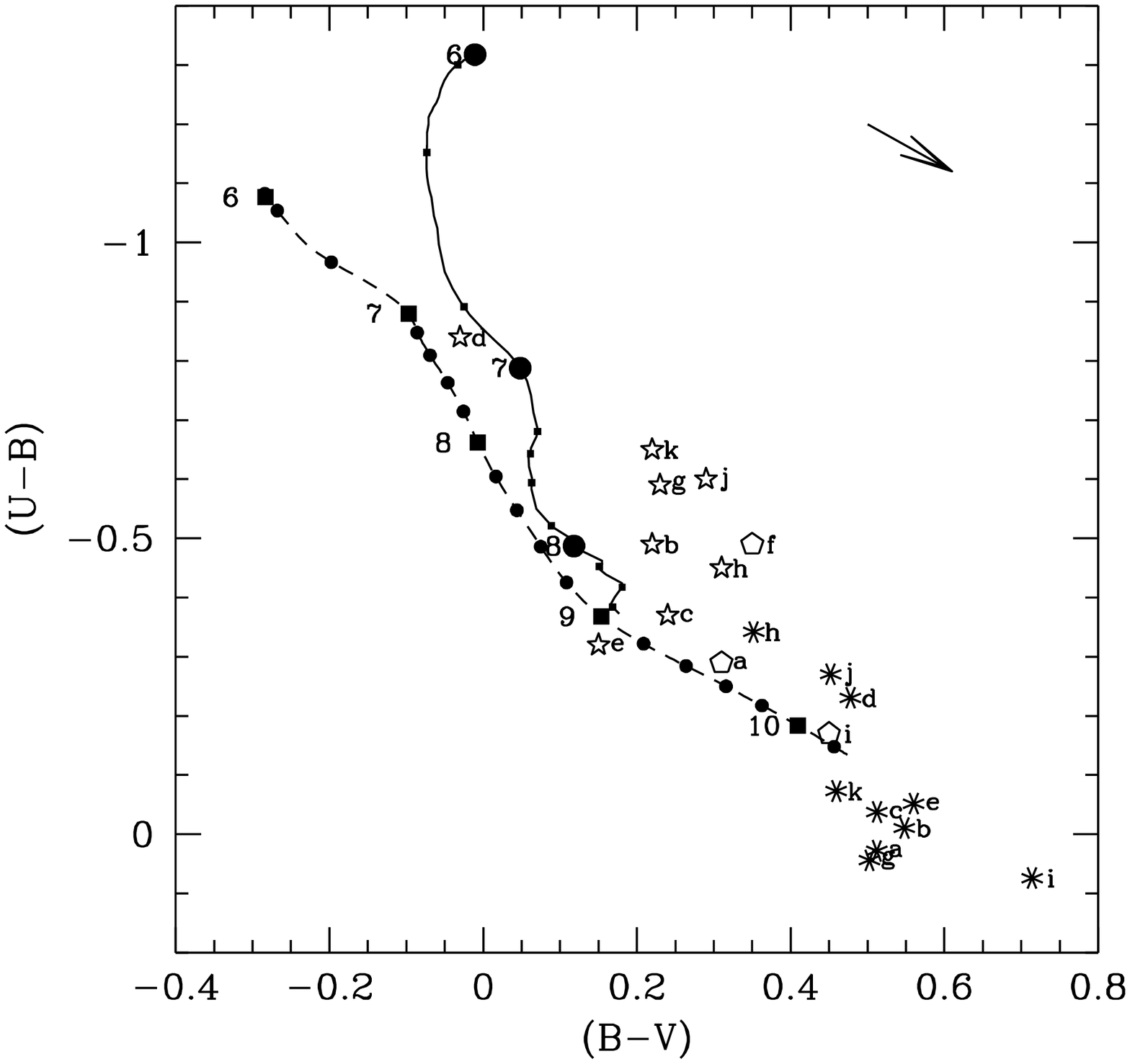,height=8.5cm}
\psfig{figure=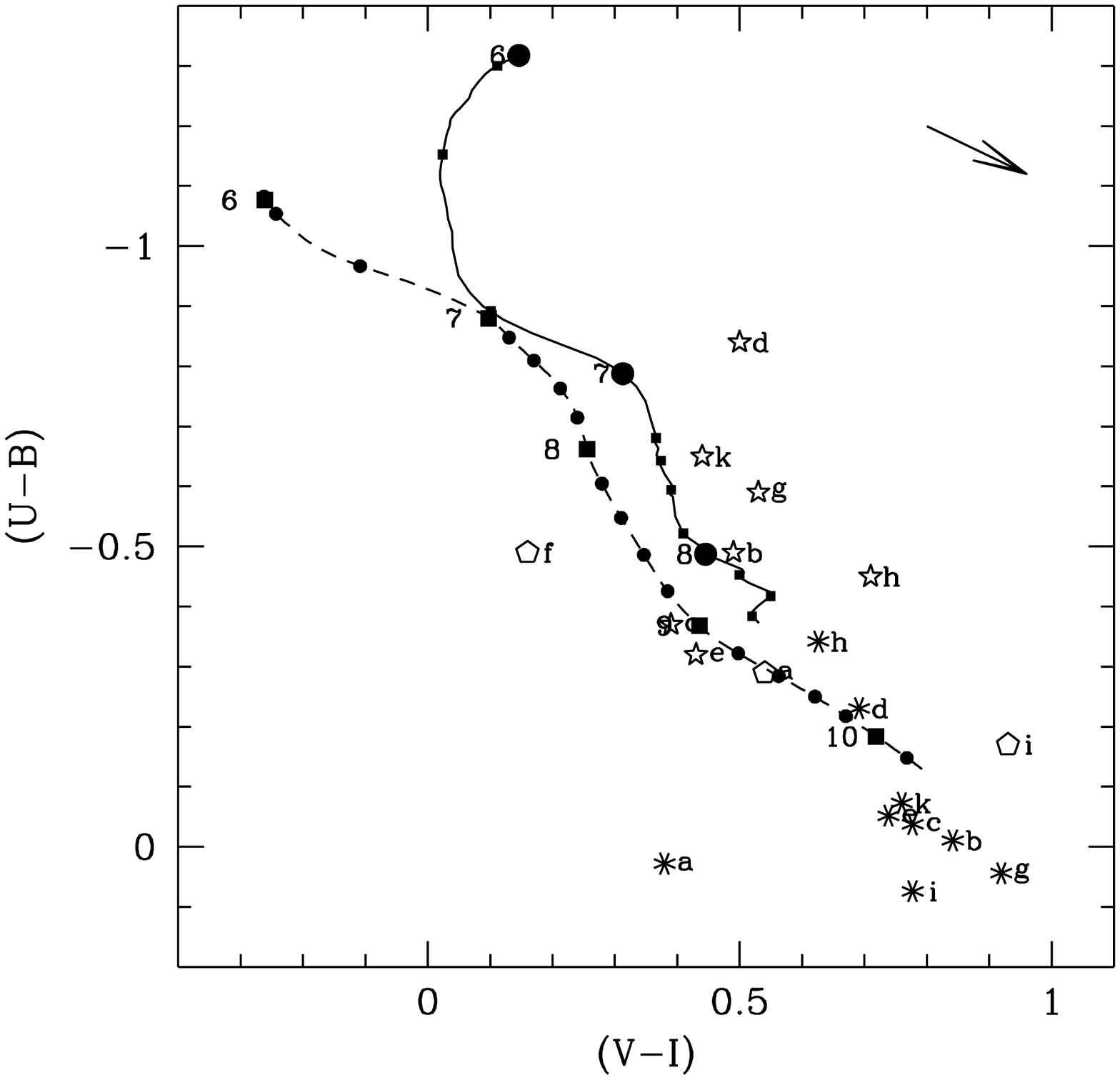,height=8.5cm}}}
\caption{Comparison of our sample with constant star formation models
in the (U--B) {\em vs.} (B--V) plane (Figure 7a) and the (U--B) {\em
vs.} (V--I) plane (Figure 7b).  The cores are plotted as open stars, the
envelopes as asterisks.  For galaxies with weak or absent cores
(NGC~2101, NGC~3955, \&\ Haro~14) the inner colors are plotted as
pentagons.  The data are compared with the constant star formation rate
models of \protect\cite{lh95} (solid line) and \protect\cite{bcmod93}
(dashed line).  The separtion of the large large symbols along the model
lines represent one dex in time, and is labeled with log(t [years]). The
smaller symbols represent 0.2 dex in time. The galaxies are labeled as
follows: (a) Haro 14, (b) NGC 625, (c) NGC 1510, (d) NGC 1705, (e) NGC
1800, (f) NGC 2101, (g) NGC 2915, (h) NGC 3125, (i) NGC 3955, (j) NGC
4670, (k) NGC 5253.  The core colors are corrected for emission line
contamination as discussed in the text. II Zw 40 and the envelope of NGC
2101 were omitted: II Zw 40 is strongly affected by emission lines and
the outer regions of NGC 2101 is badly contaminated by a nearby red
star. Also shown is the reddening vector for E(B-V)=0.1. For some
galaxies, \protect\cite{calzetal} give E(B-V) values internal to the
galaxy as derived from the Balmer decrement: NGC 1510 (0.08), NGC 1705
(0.00), NGC 1800 (0.07), NGC 3125 (0.13), and NGC 5253 (0.00).
\label{figSFHa} }
\end{figure*}

\begin{figure*}
\centerline{\hbox{\psfig{figure=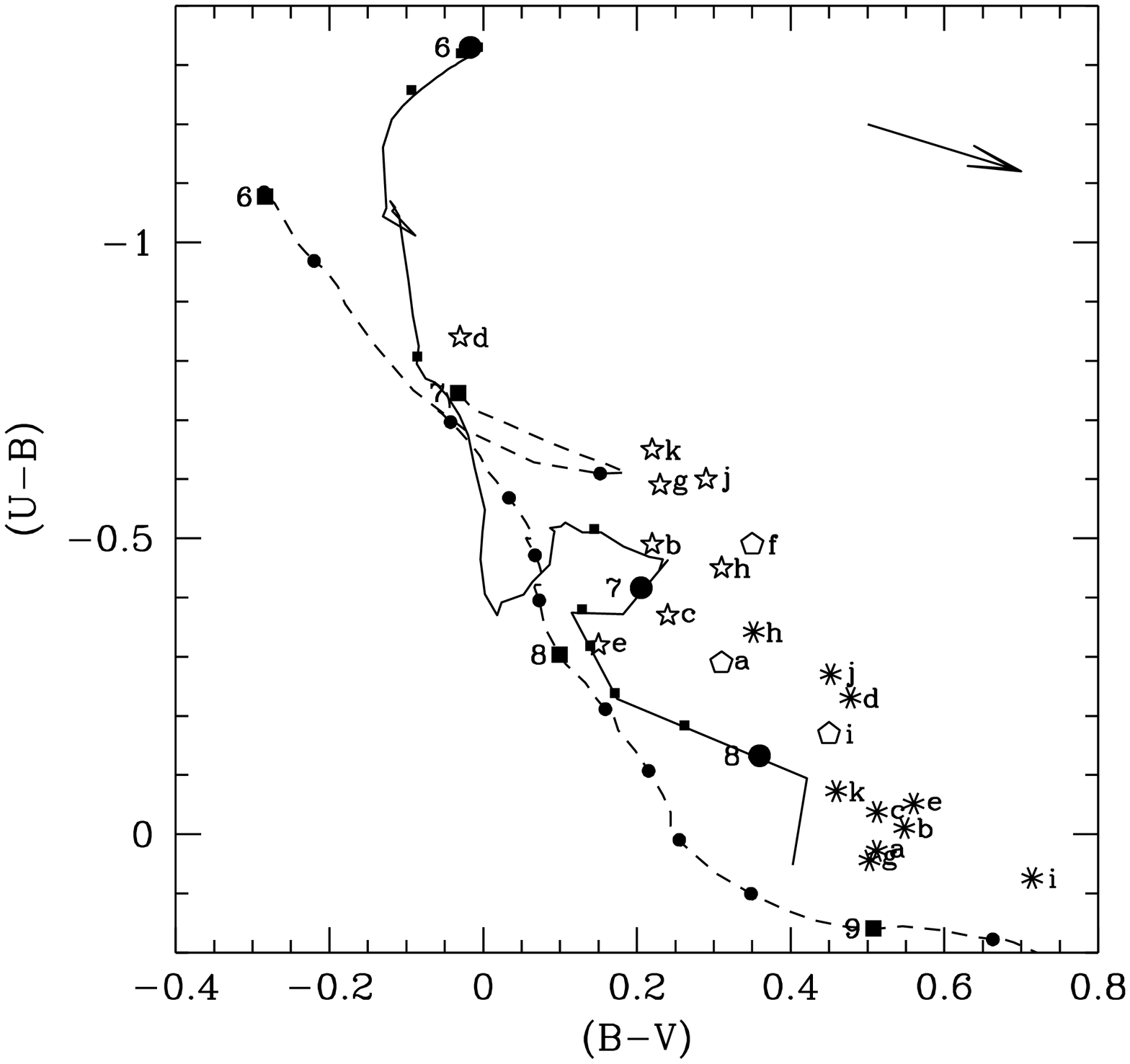,height=8.5cm}
\psfig{figure=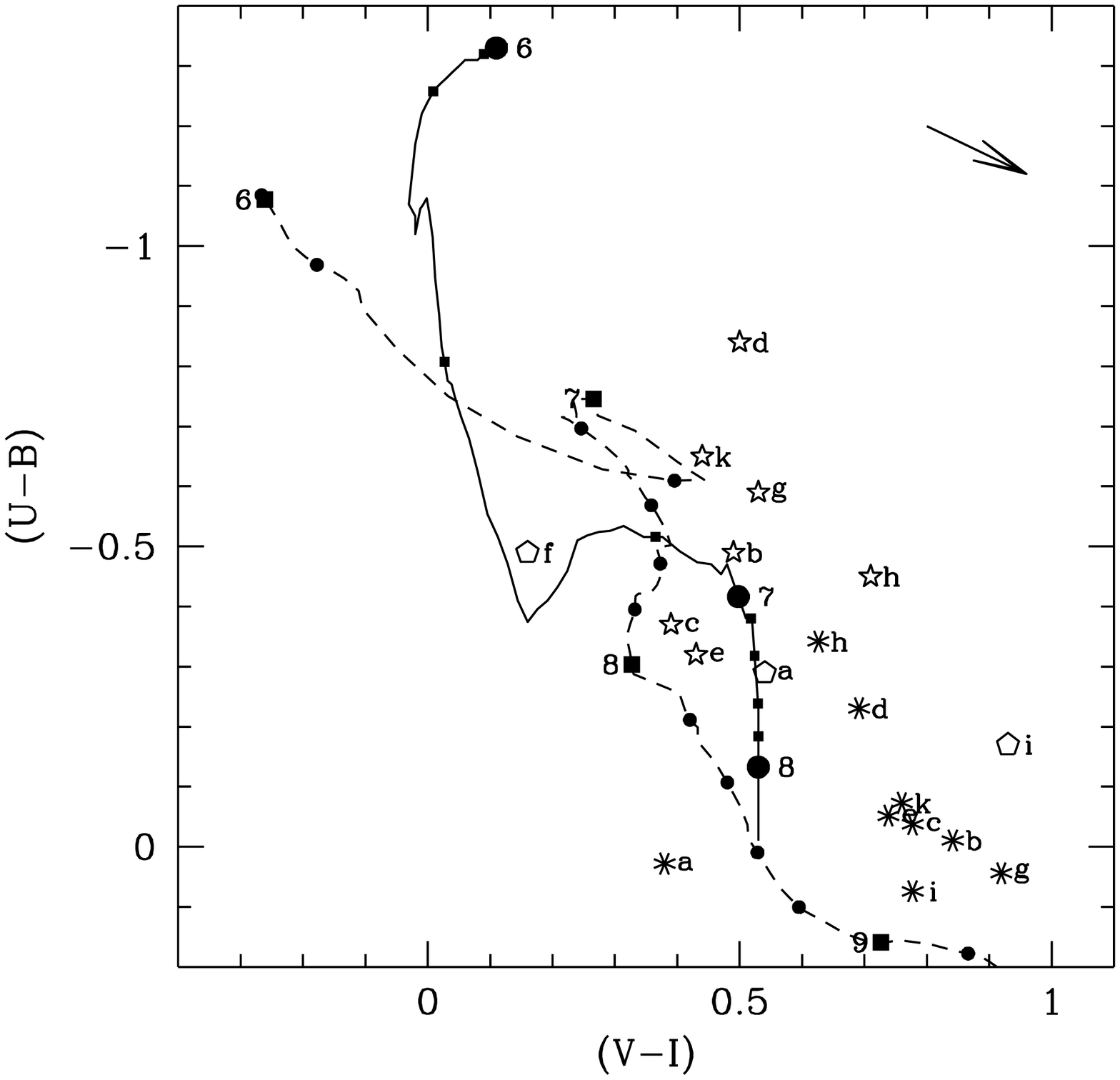,height=8.5cm}}}
\caption{Comparison with instantaneous burst models in the (U--B)
{\em vs.} (B--V) plane (Figure 8a) and the (U--B) {\em vs.} (V--I) plane
(Figure 8b).  Symbols as in Fig. \protect\ref{figSFHa}. The solid line
is the \protect\cite{lh95} model and the dashed line the
\protect\cite{bcmod93} model. The disagreement of the two models at
times $<$ 10$^7$ years is mostly due to the inclusion of the nebular
continuum in the \protect\cite{lh95} models, but not in the
\protect\cite{bcmod93} models.
\label{figSFHb}}
\end{figure*}

The comparison of data and  models is shown in Fig.\ \ref{figSFHa}
and Fig.\ \ref{figSFHb} for constant star formation and instantaneous
burst models respectively. We plot the core (star) colors and and
envelope (asterisk) colors for the sample, along with the constant
star formation models (Fig.\ \ref{figSFHa}) and the instantaneous
burst models (Fig.\ \ref{figSFHb}). Where no core subtraction could be
done, the inner ($< 0.5 r_{e,B}$) colors were used instead (pentagons).

Table \ref{tabAges} shows the ages derived from the models for an
instantaneous burst as well as for continuous star formation  based on
the $U-B$, $B-V$, and $V-I$ colors of the cores and envelopes.
To determine the ages, we find the ``distance''
\begin{equation}
d_t=\sqrt{\Sigma\left(\rm color_{gal}-color_{mod,t}\right)^2}
\end{equation}
of the observed colors to the models for each time step $t$, then
determined the age with the minimum ``distance'' for each galaxy.  The
ages and $d_t$ are given in table \ref{tabAges}. Because NGC 4670 was
assigned an arbitrary I zero point (see \cite{pap1}), we derive its age
based solely on the $U-B$, $B-V$ colors. We use the age and the absolute
blue magnitude of the population to estimate some properties of the
population, namely the total mass and the star formation rate (SFR), the
latter for constant star formation models only.


\begin{table*}
\caption{~}\label{tabAges}
\centerline{\hbox{\psfig{figure=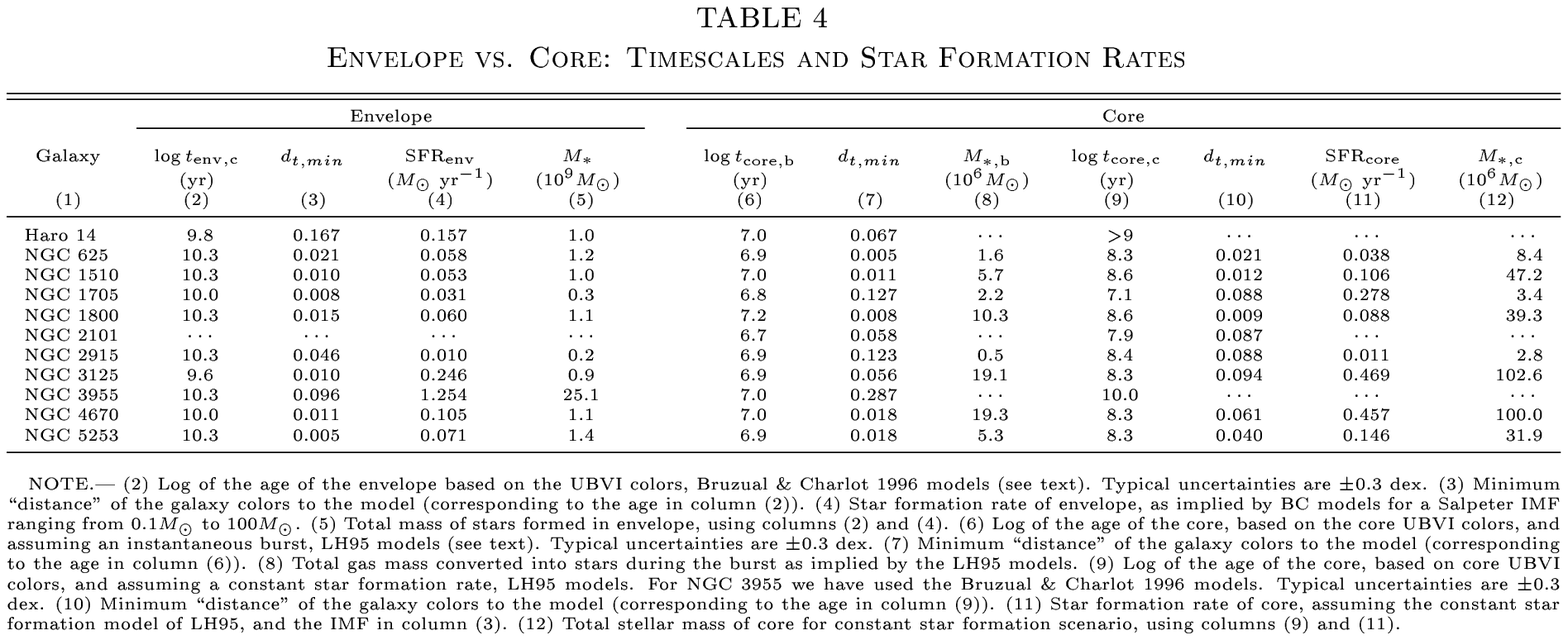}}}
\end{table*}

The models imply that while the envelopes are $\sim 10^{10}$ years old
(assuming continuous star formation), the cores are $\sim 10^7$ if they
are instantaneous bursts, or $\sim 10^8$ years old if they have been
undergoing continuous star formation. Note that NGC 3955 is an exception
to this: both its central and envelope colors are consistent with
constant star formation over a Hubble time.  This is consistent with its
spiral morphology and structure - it doesn't have a core, unlike the
majority of our sample.  Also, we note that NGC 3955 has pronounced dust
lanes and a large ratio of far-IR to \Halpha\ luminosity, so its colors
may be strongly affected by dust.

We see that the cores are much too blue to be consistent with constant
star formation over a Hubble time.  While in most cases the
instantaneous burst models provide a marginally better fit to the
broad-band data (compare $d_t$ values in Table~\ref{tabAges}), they have
a hard time reproducing the \Halpha\ fluxes.  This is because there
should be little ionizing flux at the implied burst ages of $\sim$ 10
Myr.  This is shown in Table~\ref{tabEWComp} where we give the \Halpha\
equivalent widths predicted for each core by the best fitting
\cite{lh95} models to the broad band data and compare them to the
\Halpha\ equivalent width based on \Halpha\ and broad band images of the
galaxy (see \cite{pap1} for details).  In column (4), we derive the
measured \Halpha\ equivalent width assuming that all the \Halpha\
emission is associated with the core derived from
(H$\alpha_{total}/R_{core}$, where the R band flux $R_{core}$ is
interpolated from the V and I fluxes derived for the core). Hence, here
we are assuming that the nebulae are ionization bounded and that the
core is the only, or dominant, ionizing population.  If this is not the
case, and the envelope also contributes significantly to the ionizing
flux then this \Halpha\ equivalent width can be considered an upper
limit to the true equivalent width of the core.  In column (6) we also
give the equivalent width for the entire galaxy (derived from
H$\alpha_{total}/R_{galaxy}$). Column (5) gives the equivalent widths of
\Halpha\ from the central region spectra of D.\ Calzetti (see
\S\ref{secData}). If we assume that the core is producing all or most of
the ionizing flux, then it is clear that there is better agreement with
the constant star formation models in almost all cases. Notice, however,
that the spectral equivalent width is significantly lower than the
equivalent width we get by assuming that all the \Halpha\ flux is
produced by the core for three of the galaxies: NGC 1510, NGC 1800, and
NGC 1705. While this may be due in part to the fact that the aperture of
the spectra is not an exact match to the core, it could also indicate
that a significant portion of the \Halpha\ flux is being produced
outside the core in these cases (e.g.\ NGC~1705 and NGC~1800).  However
this does not preclude our assumption that all the ionizing photons are
produced in the core (i.e.\ some \HII\ emission is far from the ionizing
source).

Most of our sample have \Halpha\ equivalent widths (measured with respect
to the core) that are more consistent with
constant star formation over $\sim 10^8$ year timescales, rather than
the best fit instantaneous burst models.  The exception is NGC~1705,
which has an equivalent width that is too low to be consistent with the
constant star formation model, but is comparable to the instantaneous
burst model. This is in large part due to the presence of NGC1705-1.  It
is a ``super'' star cluster with an age of about 13 Myr (\cite{mfdc92}),
and hence provides a strong UV-optical flux but little ionizing
continuum.  However, Meurer \etal\ (1992) show that star formation is
clearly continuing in the core which contains other smaller ionizing
clusters (cf. Meurer \etal\ 1995), and that the \Halpha\ flux is
consistent with what is expected from the stellar populations excluding
NGC1705-1. When NGC1705-1's light is removed from the core photometry,
its UBV colors are consistent with constant star formation over a $\sim$
Gyr timescale (\cite{mfdc92}).

Another indication of a strong short duration burst in a galaxy is the
presence of unusually strong Wolf-Rayet features in the spectrum. The presence
of Wolf-Rayet stars in an integrated spectrum constrains a model of an
instantaneous or nearly-instantaneous burst to an age between $3-7$ Myr
(cf. \cite{lh95}). Four of the galaxies in our sample (NGC 1510, II Zw
40, NGC 3125, and NGC 5253) have broad emission lines at 4686\AA\ due to
HeII (\cite{Conti91} and sources therein), usually the strongest line at
optical wavelengths for Wolf-Rayet stars. 

However, a recent burst that produced the Wolf-Rayet stars in these galaxies
may not account for their
entire cores.  A case in point is NGC~5253, probably the best studied
``Wolf-Rayet galaxy'' in our sample.  Schaerer \etal\
(1997\markcite{SCKM97}) find that the Wolf-Rayet features are localized
to two knots, i.e.\ clusters, and derive ages of 2.8, and 4.3 Myr for
them.  Calzetti \etal, (1997\markcite{C97}) studied the stellar
populations in NGC~5253 at high resolution using broad and narrow band
images taken with the Hubble Space Telescope.  They confirm the ages of
the Wolf-Rayet clusters, but show that other prominent clusters have
ages up to $\sim 50$ Myr, and moreover, that the colors of the diffusely
distributed stars between the clusters is consistent with continuous
star formation for $\sim 500$ Myr timescales.  This is fairly consistent
with the 200 Myr duration we derive for the core (combined clusters and
diffusely distributed stars) in the continuous star formation model.

The overall picture we find is that the star formation histories of the
``starburst'' cores is complex, most likely consisting of continuous
star formation, interspersed with strong enhancements as clusters form.
While the timescale for cluster formation is probably $\leq$ few Myr, as
implied by (localized) Wolf-Rayet features in some cases, the timescale
for the whole of the core to form is less clear.  At a minimum it should
be the luminosity weighted mean age of $\sim 10$ Myr derived from
instantaneous burst models.  As discussed above, however, timescales 
of $\sim 100$ Myr are more likely for the total core.

\section{Evolutionary connections}
\label{secHost}

Starbursting dwarfs are similar to dE and dI galaxies in having
exponential surface brightness profiles, at least at large radii.  If
there are evolutionary connections between starburst dwarfs and the
other types, then it is the envelope that tells us about the host, hence
the progenitor and successor to the current starburst system. In Figure
\ref{figDiskProps} and Fig.~\ref{figDiskProps2}, we compare the envelope
structural properties of various samples of dwarfs. The dI sample is
taken from Patterson \&\ Thuan (1996, hereafter \cite{pt96}).  For the
dEs, we chose the \cite{cb87} dwarf elliptical sample over the more
extensive \cite{bc93} sample so as to compare results with exponential
fits to elliptical annular surface brightness profiles where possible.

Figure \ref{figDiskProps} shows a clear structural difference in the
envelopes of starburst dwarfs compared to other dwarf galaxies.  This is
clear in both the $\alpha^{-1}$, $\mu_{0,c}$ plane and in the histograms
of the two quantities.  In terms of average quantities
$\langle\log(\alpha^{-1} [{\rm kpc}]) \rangle$, $\langle \mu_{0,c}
\rangle$, starburst dwarfs have -0.1, 21.3 (combined all three samples);
dIs have 0.2, 23.6; and dEs have -0.3, 22.9.  The median
($\log(\alpha^{-1} [{\rm kpc}]) $, $\mu_{0,c} $) quantities are -0.2,
21.2 for starburst dwarfs; 0.2, 23.7 for dIs; and -0.3, 22.8 for dEs.
Considering the differences in their exponential envelopes, what
evolutionary connections are allowed between dwarf galaxy classes?  We
now consider this question.

\subsection{Progenitors of starburst dwarfs}

The progenitors to starburst dwarfs must be gas rich systems.  This
rules out dEs, which are gas poor, as the progenitors.  Dwarf irregulars
have commonly been thought to be the most likely candidate to the
preburst state (e.g.\ \cite{dp88}).  However, the typical BCD has an
envelope with scale length about a factor of two smaller than typical
dIs, and central surface brightness about a factor of ten higher than
typical dIs.  Much of the difference in scale lengths may be attributed
to selection effects.  Both our sample and the P96 sample have a (loose)
absolute magnitude limit ($M_B \gtrsim -18$), as does the PT sample of
dI galaxies.  This limiting magnitude (for a pure exponential profile)
is shown as a dotted line in Fig.~\ref{figDiskProps}.  Galaxies
significantly above this line will not be selected except perhaps in the
T97 sample; i.e.\ there is a bias against large scale length, high
surface brightness galaxies -- they are not dwarfs. However, such
galaxies do exist, and occasionally do make it into ``dwarf'' samples.

However, there is still the factor of $\sim$ ten difference in central
surface brightness to explain.  Figure~\ref{figDiskProps} suggests a
dividing line at $\mu_{0,c} \approx 22\, {\rm mag\, arcsec^{-2}}$;
envelopes of starburst dwarfs relatively rarely have (extrapolated)
central intensities less than this, while dIs rarely have higher central
intensities. This too could in principle be due in part to a sample
selection effect: do the galaxy catalogs that are used to generate
samples of dIs (e.g. the UGC) sytematically exclude high surface
brightness irregular galaxies from the list of objects classified as
dIs? Is there are population of non-bursting, gas-rich, dwarfs with high
central surface brightnesses and short scale lengths? The qualitative
answer is ``yes'', since our sample includes objects like Haro~14,
NGC~625, and NGC~2101 (none of which have strong blue starbursting
cores).  While it is beyond the scope of the present paper to quantify
the importance of this effect, it is important to keep this caveat in
mind.

Selection effects aside, could a measurement bias cause the segregation
of types in Fig.~\ref{figDiskProps}?  The concern is that a low surface
brightness envelope would be difficult to detect, with the available
photometry, if a strong burst were present. This is unlikely to be a
significant effect since neither our sample nor those of P96 nor T97
contain `bare cores' (i.e. an envelope is always detected around the
bright blue starburst core).  Nevertheless, we have performed some
simple simulations to test this possible measurement effect.  We assume
a double exponential structure for burst plus envelope.  We take
$\alpha^{-1} = 120$ pc, $\mu_0 = 18.5\, {\rm mag\, arcsec^{-2}}$ for the
burst, and vary the envelope structural parameters.  We find that our
photometry, which typically reaches $\mu(B) \approx 24.5\, {\rm mag\,
arcsec^{-2}}$ should have no problem detecting the exponential profile
of an envelope with $\mu_{0,c} = 22\, {\rm mag\, arcsec^{-2}}$ and
$\alpha^{-1}$ in the range of 0.5 to 2 kpc.  However at $\mu_{0,c} =
23\, {\rm mag\, arcsec^{-2}}$, it would be difficult to detect short
scale length, $\alpha^{-1} \lesssim 0.5$ kpc, (face-on) envelopes with
our photometry.  Envelopes with $\mu_{0,c} \gtrsim 24\, {\rm mag\,
arcsec^{-2}}$ would be difficult to detect, unless at a high
inclination.  The photometry of T97 frequently goes about a magnitude
deeper (albeit in the V band).  Consequently they should be able to
detect the exponential envelopes with $\mu_{0,c}(V) = 24\, {\rm mag\,
arcsec^{-2}}$ and $\alpha^{-1} \gtrsim 1$ kpc.  The photometry of P96 is
the deepest, typically reaching $\mu(B) \approx 26.5\, {\rm mag\,
arcsec^{-2}}$.  They should be able to detect exponential envelopes
$\mu_{0,c}(B) = 24\, {\rm mag\, arcsec^{-2}}$ and $\alpha^{-1} \gtrsim
0.7$ kpc.  Our conclusion is that enough of the envelope parameter space
is available to our study or others that moderately low surface
brightness envelopes should have been detected if they exist.  Deeper
observations and more detailed simulations would be useful for limiting
and characterizing the envelope measurement biases.

We are left with the result that there is a real difference between
the envelope structure of starburst dwarfs and dwarf irregulars.  Could
this be due to a dynamical response of the envelope to the starburst
core?  If a dI where to have a central burst, gas will have to accrete
into the center.  The response of the envelope to the increased
potential will be a contraction, and consequently an increased surface
brightness.  Conversely, a wind associated with the starburst will
expel mass from the center, and the stellar distribution will expand
and fade in surface brightness in response.  For a fractional change of
mass $f$, the expansion/contraction factor $e \approx (1-f)/(1-2f)$ if
the mass loss/gain occurs on a timescale much less than the dynamical
time, and $e \approx 1/(1-f)$ if the mass loss/gain occurs on a
timescale much greater than the dynamical time (\cite{ya87}).
Considering the fairly long core star formation timescales we derive, it
is likely that the mass loss or gain is slow.  

The problem is that for evolution between dIs and starburst dwarfs we
need $e \approx 3$, or $f \approx 2/3$.  This huge change in mass does
not seem feasible.  First, consider the effects of winds.  In Marlowe
\etal\ (1995), we discuss in detail the dynamics of those galaxies
showing evidence of large scale outflows. In that paper, we roughly
estimate the energy output of the starbursting region based on
galaxy-wide parameters. With the improved population models for the
cores, we can now refine our estimates for the mechanical energy output
of the starbursts in those galaxies with a windy \Halpha\
morphology. The new estimates are shown in Table \ref{tabMechLum}, which
also includes the estimates for those galaxies that do not have any
evidence of large scale outflows.  The main conclusions of Marlowe
\etal\ (1995) are not changed significantly by these new energy
estimates: the mechanical luminosities of the present bursts are
insufficient to remove large amounts of gas from the galaxies in our
sample, but {\it are} sufficient to drive the large-scale expansion seen
in the H$\alpha$ bubbles and to allow much/most of the 
newly-created metals to be ejected from the galaxy:
adopting the termin0logy of \cite{dYH94} and \cite{mf99}, the bursts
can accomplish ``blow out'', but not ``blow away''.
Second, it should be remembered that dwarf galaxies are dark matter
dominated.  Both dI and starburst dwarfs are known to have dark matter
halos dominating even into the optical face of the galaxy (e.g.\ DDO
154: \cite{cb89}; NGC~2915: \cite{mcbf96}).  Hence even if all the ISM
were accreted or removed, there is likely to be little effect on the
potential well depth, and hence little expansion or contraction.

\begin{deluxetable}{lrrccccc}
\small
\tablecaption{Star formation and Energy Rates of the Starbursts
\label{tabMechLum}
}
\tablehead{
\colhead{Galaxy} & \colhead{$L_{\rm Bol,B}$} &
\colhead{$L_{\rm Bol,C}$} & \colhead{$M_{*,{\rm B}}$} &
\colhead{SFR$_C$}  & \colhead{SFR$_{LyC}$} & \colhead{$\log\dot{E}_{\rm B}$}  &
\colhead{$\log\dot{E}_{\rm C}$}  \\
\colhead{} & \colhead{($10^8 L_\odot$)} &
\colhead{($10^8 L_\odot$)} & \colhead{($10^6 M_\odot$)} &
\colhead{($M_\odot$ yr$^{-1}$)} &\colhead{($M_\odot$ yr$^{-1}$)} &
\colhead{(ergs s$^{-1}$)} &
\colhead{(ergs s$^{-1}$)} \\
\colhead{(1)} & \colhead{(2)} &
\colhead{(3)} & \colhead{(4)} &
\colhead{(5)} & \colhead{(6)} &
\colhead{(7)} & \colhead{(8)}
}
\startdata
 NGC 625   &3.1     & 3.2      & 0.61    & 0.038  & 0.040 & 40.1    & 40.0    \nl
 NGC 1510  &9.3     & 9.3      & 2.24    & 0.106  & 0.068 &40.7    & 40.5    \nl
 NGC 1705  &6.7     & 14.0     & 0.87    & 0.278  & 0.04\phn & 40.4    & 40.4    \nl
 NGC 1800  &7.7     & 7.7      & 4.05    & 0.088  & 0.058& 40.9    & 40.4    \nl
 NGC 2915  &0.9     & 0.9      & 0.18    & 0.011  & 0.007 & 39.5    & 39.5    \nl
 NGC 3125  &37.5    & 39.3     & 7.47    & 0.469  & 0.67\phn  & 41.2    & 41.1    \nl
 NGC 4670  &38.4    & 38.3     & 7.56    & 0.457  & 0.34\phn & 41.2    & 41.1    \nl
 NGC 5253  &13.2    & 12.2     & 2.08    & 0.146  & 0.17\phn & 40.6    & 40.6    \nl
\enddata
\tablecomments{Values in this table (except Col. (6)) are 
derived from the LH95 
models using the core blue luminosity
(Table \protect\ref{tabDCProp} 
and the ages as derived in Table \protect\ref{tabAges}).
(2) Bolometric luminosity of the galaxy core, assuming
the instantaneous burst models of LH95 as described in
\S\protect{\ref{secCETimes}}.
(3) Bolometric luminosity of the galaxy core, assuming the constant
star formation model of LH95, as described in
\S\ref{secCETimes}. 
(4) Mass of gas converted into stars in the core,
assuming LH95 instanteous burst models. (5) Star formation
rate in the core, assuming LH95 constant star formation rate star formation
models and IMF described in \S3.3. (6) Star formation
rate calculated from the same metallicity and IMF as in (5), but
using the Lyman continuum as calculated from the H$\alpha$
luminosities in Paper I (Table 7) assuming case B conditions
at $T=10^4$ K (Osterbrock 1989). (7) Rate of energy deposited by the
core, assuming LH95 instantaeous burst models. (8) Rate of energy
deposited by the core, assuming LH95 constant star
formation rate models. }
\end{deluxetable}

\begin{figure*}
\centerline{\hbox{\psfig{figure=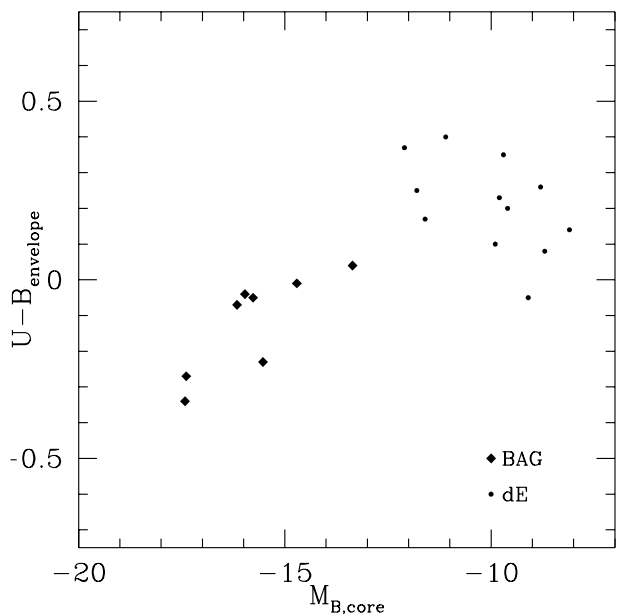,height=8.5cm}}}
\caption{Envelope U--B color {\em vs.\/} core absolute blue
magnitude. Blue amorphous galaxies (BAG): this sample (diamonds), dEs:
\protect\cite{cb87} (circles).
\label{figCorr}}
\end{figure*}

Another way to enhance the $\mu_{0,c}$ of the envelope is to have galaxy
wide star formation correlated with the core formation.  A hint of this
is suggested in our sample as can be seen in Figure \ref{figCorr} which
plots the envelope U-B colors versus the absolute blue magnitude of the
cores: brighter cores are correlated with bluer envelopes. This suggests
that whatever event triggered the intense star formation in the core may
also have triggered increased star formation throughout the galaxy on a
less intense scale.  However, there are three inadequacies with this idea.
First, the color profiles show the envelopes to be redder, and hence
have a longer star formation timescale (in most cases roughly a Hubble
time) than the core.  Second, we still have the problem with scale: we
need a factor of about ten enhancement in surface brightness.  This
seems excessively large considering the rather moderate contribution of
the core. Third, there may be more contamination of the envelope by the core
(starburst) light in the larger systems.

We are left with the likelihood that there is little evolution between dI
and starburst dwarf morphologies.  Furthermore, high surface brightness
starbursts, as we know them, do not occur in low surface brightness
($\mu_{0,c} > 22\, {\rm mag\, arcsec^{-2}}$) host galaxies. Rather they
require a relatively high surface brightness host ($\mu_{0,c} < 22\, {\rm
mag\, arcsec^{-2}}$).

What then are the progenitors of starbursting dwarfs?  They must have
moderately high central surface brightness and relatively short scale
length.  The most likely contenders, we believe, would have pure
exponential or plateau profiles. Some of these objects are found amongst
samples of dIs, and comprise the dIs with the highest surface-brightness
(see Figure 5). Other progenitors may already have one of the various
`starburst' dwarf designations.  However, the `starburst' classification
would be incorrect in these cases. It would be based mainly on high
surface brightness (relative to typical dIs). Closer examination would
show these galaxies to have modest emission line equivalent widths,
small or absent color gradients between the central regions and the
outer extremes, and colors throughout that are consistent with a current
star-formation rate not too different from the past average over a
Hubble time.

This idea is similar to the scenario of Meurer \etal\
(1994\markcite{mmc94}): at the peak of its star formation burst the
system would look like a strong core/envelope system such as NGC~1705,
while before or after the burst it would like a weak core system such as
Haro 14, NGC~625, or NGC~2101. That is, we would expect the class of
amorphous galaxies to contain systems in both the `burst' (strong blue
core) and `inter-burst' (weak/absent core) states. Such galaxies might
cycle between these two states intermittently until the gas is used up
or expelled (see below). Since the starburst cores seem to involve only
a few percent at most of the total galaxy stellar mass, there is no
strong constraint on the duty cycle (e.g. the galaxy could go through
many such weak bursts and spend much of its time in this phase). Salzer
\etal\ (1995) \markcite{salzetal95} presents a large survey of bursting
(including weakly bursting galaxies) and ``normal'' dwarfs. The
fractional number density of bursting dwarfs is about 1/3. This suggests
that starburst dwarfs at $M_B \approx -16$ are ``on'' about 1/3 of the
time. With a typical burst duration of $\sim 2 \times 10^8$ yr, the
typical quiescent period would last $\sim 6 \times 10^8$ years. Thus,
over 10 Gyr, a galaxy could conceivably burst about 10 times, with a
total star forming time of a few $10^9$ years. This is on the order of
the gas consumption times seen in most blue amorphous galaxies (cf.
\cite{mhws95}).

About 1/3 of the dIs studied by \cite{pt96} also have core/envelope
structures, but the central intensity of even the {\em core\/} is much
lower - typically $\mu_{0,c}(B) \gtrsim 21\, {\rm mag\, arcsec^{-2}}$.
This suggests that dIs also have episodes of star formation.  However
the intensity of star formation is generally lower, and apparently does not
usually produce the characteristics (high surface brightness in continuum and
lines) which we normally associate with starbursts.  

There may be a good dynamical basis for a correlation between burst
intensity and host structure.  There is a correlation between dark
matter halo central density $\rho_0$ and galaxy surface brightness - low
surface brightness dIs have low $\rho_0$ (\cite{dbm97}) while high
surface brightness starburst dwarfs have high $\rho_0$ (\cite{mcbf96};
\cite{mssk98}). In both cases the DM dominates, even into the centers of
the galaxies.  If efficient star formation, i.e. a burst of star
formation, occurs when a disk becomes unstable to self gravitation and
forms stars on the shortest possible timescale - the dynamical timescale,
then we expect the surface brightness of the starburst $\propto \rho_0$
(\cite{mhlll97}).  A (weak) correlation in this sense is also observed
for starbursts in normal systems (\cite{mhlll97}).

\subsection{Evolutionary endpoints of starburst dwarfs}

In the previous section, we have explored the idea that the progenitors
of the dwarf starburst systems are most likely the ``coreless''
amorphous galaxies like Haro~14, NGC~625, and NGC~2101 in the present
sample and dIs of the highest surface brightness. 
We have also suggested that such galaxies may cycle repeatedly
between ``on'' and ``off'' states (a recurring series of mild starbursts).
Thus, on intermediate time-scales the evolutionary descendants of
starburst dwarfs are likely to be objects like the three listed above.
In this section we explore the plausibility of a more drastic and
long-term evolutionary path which might occur if a starburst dwarf
were able to consume or lose its gas supply. We first consider what
the evolutionary path would be {\it if} one of our bursting dwarfs
were to lose its gas and cease forming stars. We then consider whether
this scenario is physically plausible.

Dwarf ellipticals are the most likely candidates for the 
ultimate end product of a starbursting dwarf that loses all its gas.
This possible evolutionary link has long been
proposed (e.g.\ \cite{bmcm86}; \cite{dh91}, \cite{mfdc92}).  The
envelopes currently seen in most blue amorphous galaxies are too bright
and too blue to be dE envelopes (however cf.\ NGC~2915).  Starburst
dwarf envelopes are consistent with constant star formation for several
Gyr.  If this star formation were turned off, however, after another
several Gyr, the blue amorphous galaxy envelopes would fade
several magnitudes and become redder. For example, a standard Bruzual \&
Charlot (1996) model (Salpeter IMF from 0.1 to 125 M$_{\odot}$ with 40\%
solar metallicity) for star-formation with a duration of 3 to 10 Gyr
will fade by by about 2.4 magnitudes in the next 10 Gyr if
star-formation is shut off.  This would bring starburst dwarf envelopes
well into the $\mu_{0,c}$ regime of dEs (Fig \ref{figDiskProps} and
Fig.~\ref{figDiskProps2}). 

\begin{figure*}
\centerline{\hbox{\psfig{figure=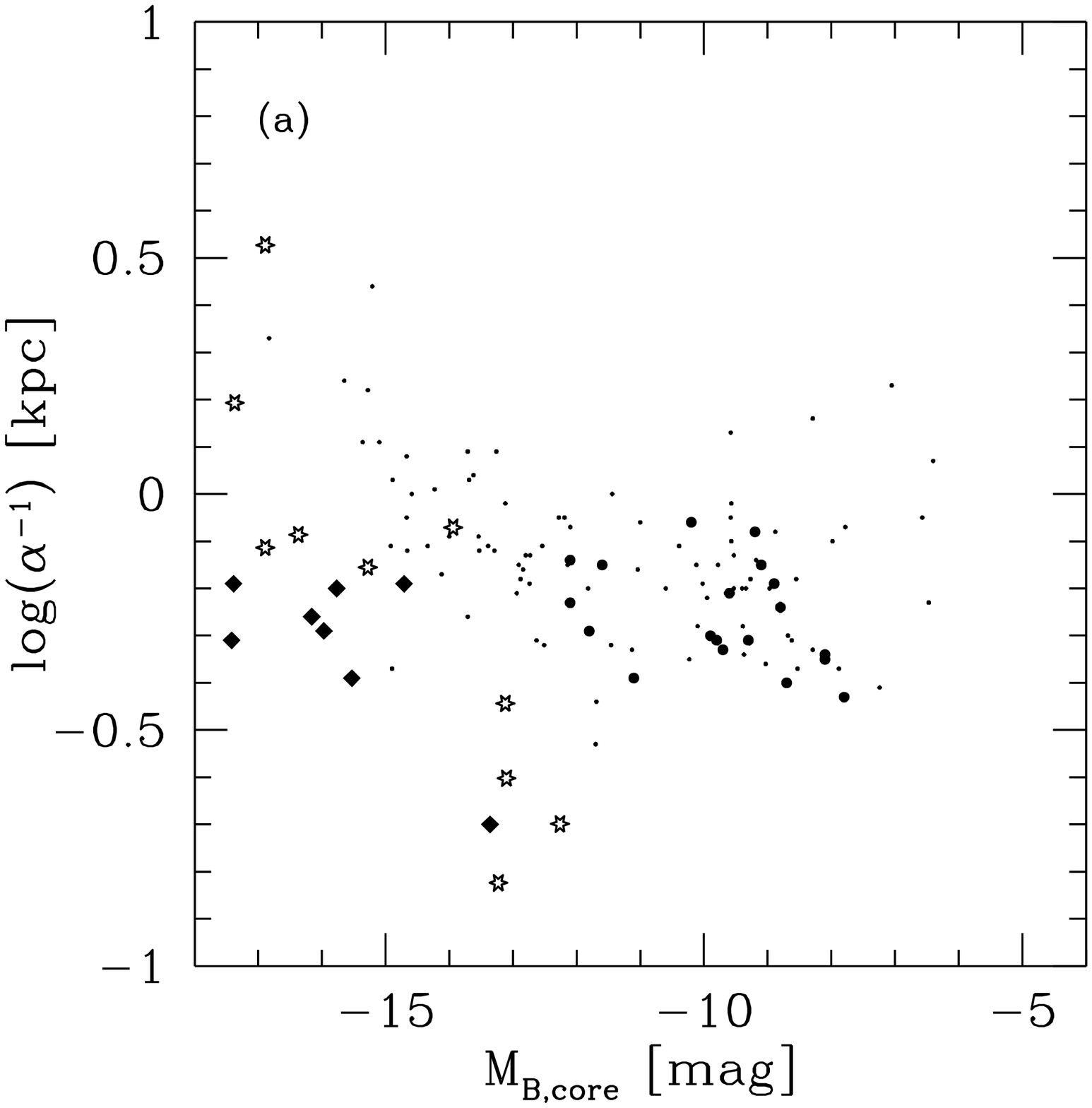,height=8.5cm}
\psfig{figure=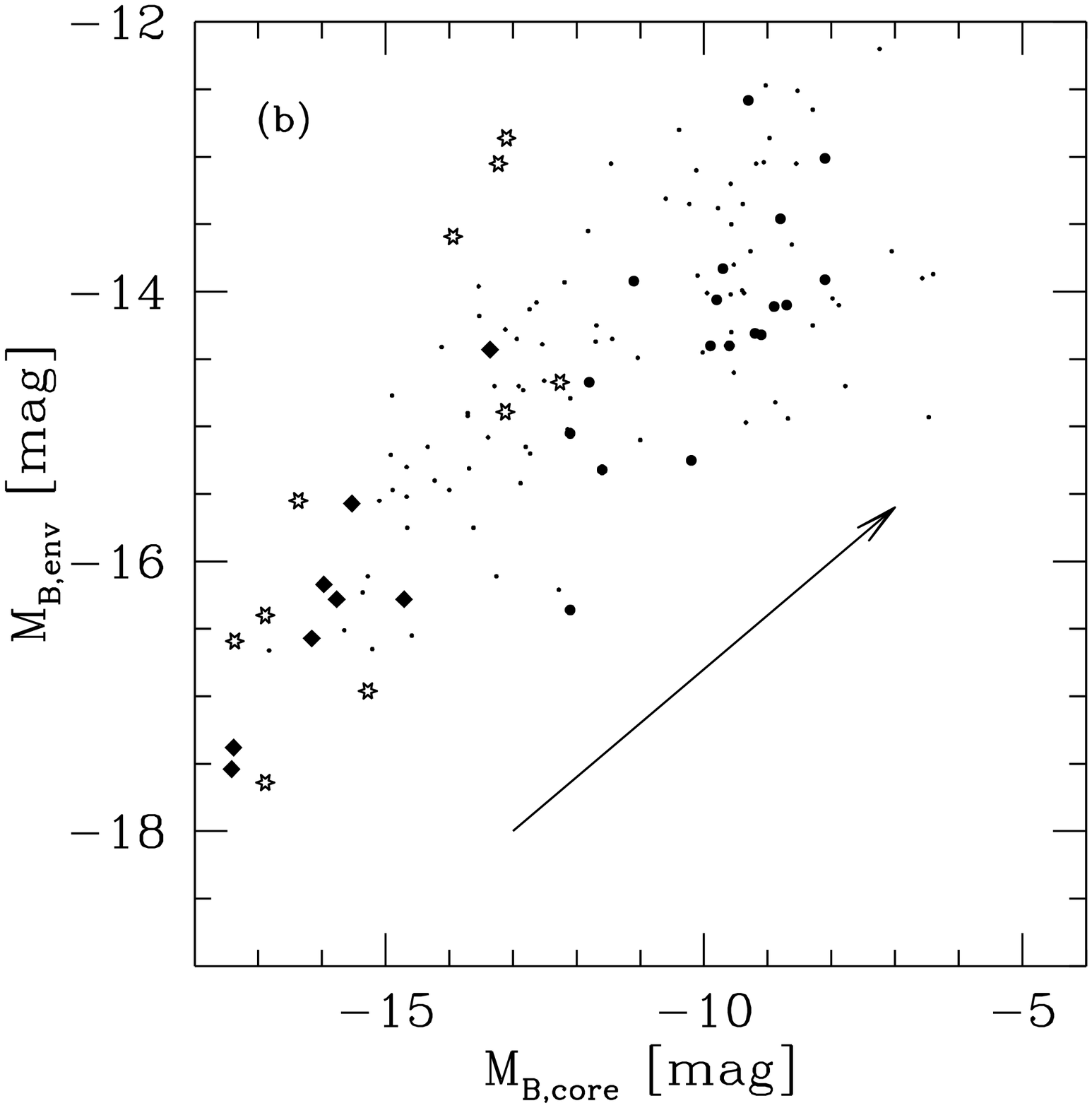,height=8.5cm}}}
\caption{Comparison of dwarf ellipticals with cores to starbursting
dwarfs.  (a) Scale length vs. core absolute blue magnitude. Starbursting
dwarfs: this sample (diamonds), P96 sample (stars), and dEs:
\protect\cite{cb87} (large circles) and \protect\cite{bc93} (small
circles). (b) M$_{B,core}$ vs. M$_{B,env}$ (symbols as in (a)). The
vector shows the effect of cutting off star-formation in both the core
and the envelope and allowing both to fade for 10 Gyr. The core fades
more than the envelope because it has had a much shorter duration of
star-formation (see text for details.)
\label{figdEcores}}
\end{figure*}

A fraction of the dEs in the Caldwell \&\ Bothun
(1987\markcite{cb87}) sample have cores, and are designated as
``nucleated'' dEs. We plot the scale length  versus
$M_{B,core}$ and $M_{B,core}$ versus $M_{B,env}$ for these galaxies
along with starburst dwarfs in Figure
\ref{figdEcores}. There is more scatter for dEs, which is what one
might expect if they are faded compact galaxies: the scale length
would not change drastically, though the core magnitude would. The
apparent lower limit to the scale length for dEs probably
corresponds to a selection effect and seeing limitation since all
these galaxies are in the Virgo and Fornax clusters.

If the cores of the amorphous galaxies and BCDs fade enough, the faded
galaxies will move into the area occupied by these nucleated dEs.  The
\cite{bcmod93} models predict that after 10 Gyr, a 0.1 Gyr duration
burst will fade about 6 magnitudes. Thus, if we were to shut off the
star-formation for 10 Gyr in
both the core and envelope of one of these starburst
dwarfs, the stronger differential fading of the core (6 magnitudes)
vs. the envelope (2.4 magnitudes) would move the galaxy along the
core vs. envelope sequence in
Figure \ref{figdEcores}b from the domain of the starburst dwarfs into
the domain of the nucleated dEs. However, the fading of the core will not be
so dramatic if cores
result from repeated weak bursts rather than a single strong burst
(the limiting case of many weak bursts will approximate a constant
star-formation rate, which would then fade by only 2.4 magnitudes
in 10 Gyr - see above).

The major problem with starburst dwarfs evolving into dEs is gas removal. As
previously noted, the mechanical luminosity of the cores is
insufficient to remove all the ISM; at least not over the core star
formation timescale. There is also the problem that dwarf ellipticals
are found predominantly in clusters (\cite{vs91}), while starburst
dwarfs are more numerous in the field (\cite{sal89}).  One way around
these problems is to have multiple bursts.  The ISM consumed in star
formation and ejected in winds would then be multiplied by the number of
bursts.  In clusters, evolution would occur at a more rapid pace as
there are more frequent perturbing ``triggers''. In addition, ram
pressure stripping of the ISM will play a more important role in
clusters.  This might explain the environmental differences between dEs
and starburst dwarfs.
 
The multiple burst scenario can also explain the correlation seen in dEs
between exponential envelope color and the core strength shown in
Fig.~\ref{figCorr}.  This correlation is in the sense that brighter
cores have redder envelopes, opposite in sense to what is seen for
starburst dwarfs.  \markcite{cb87}Caldwell \&\ Bothun (1987) attribute
this to metallicity effects: dEs that have undergone multiple bursts
before fading have accumulated more metals in the galaxy than those with
fewer bursts. Multiple bursts would also build up the core strength as
low mass stars from previous bursts would continue to contribute light
to the core. It may not be possible to have such a correlation from
cores formed in a single burst, since this would involve a much stronger
burst, one more likely to expel a metal enriched ISM in a wind.

We conclude that while it is possible in a photometric or purely
phenomenological sense for the evolutionary endpoint of starburst dwarfs
to be dEs, there is no compelling reason to expect that {\it typical
starburst
dwarfs in the present cosmological epoch} (such as those we have studied
in this paper) will follow this pathway any time soon. In fact, we have
argued that these starbursting dwarfs are experiencing rather mild recurring
events, so that the depletion/loss of gas (and the resulting cessation
of star-formation) will occur slowly.
Turning this around, since the progenitors of present-day dEs
were star-forming gas-rich dwarfs that succeeded in consuming/ejecting/losing
their ISM, these progenitors must have
undergone bursts of star-formation that were considerably more intense
than the bursts seen in typical BCDs today.

\section{Conclusions}\label{secConc}

Our analysis of blue amorphous galaxies shows that their surface
brightness and color profiles typically consist of a young burst
occurring in an older, enveloping population (\cite{pap1}). The
underlying galaxy surface brightness profile (which dominates beyond
about 1.5 times the half-light radius) is consistent with an exponential
profile. The young burst can be characterized by a central excess (the
core) of this exponential profile.  Some objects loosely classified as
starburst dwarfs (our sample and others) instead have a purely
exponential or central plateau structure, instead of the more common
core - envelope structure.  In these ``coreless'' objects there is
little if any color gradient - the colors throughout are consistent with
a roughly constant star-formation rate over the last few Gyr.

We show that the underlying galaxies in our sample have colors
consistent with roughly constant star formation over a Hubble time. The
cores are younger and have broadband colors consistent with ages of
$\sim 10$ Myr if formed in an instantaneous burst or $\sim 100$ Myr if
they result from a constant rate star formation. Most cores have
\Halpha\ equivalent widths more consistent with the constant star
formation rate models.  The core typically provides about half the light
of the galaxy, and comprises a few percent of the stellar mass, as
derived from our population models (Table \ref{tabAges}) We thus
conclude that the burst is a relatively modest event in the life of the
underlying galaxy.  These bursts are unlike the short duration dominant
bursts hypothesized to occur in dwarf galaxies (\cite{bf96}) in order to
explain the faint blue galaxy problem.

The properties of our sample of predominantly morphologically selected
dwarf starbursts has a high degree of overlap with those of other dwarf
starburst samples, i.e.\ those going by the monikers BCD and \HII\
galaxies.  They have similar core colors, envelope colors, \Halpha\
equivalent widths, core to envelope flux ratios, and envelope structural
parameters.  Hence we conclude that all these starburst dwarf samples
represent the same basic physical phenomenon: recent and intense high
mass star formation occurring in a redder and presumably older host.

Our study and the similar studies of Telles \etal\ (1997a,b) and
Papaderos \etal\ (1996a,b) show that the structure of the envelope is
different from the exponential profiles of most dI galaxies: starburst dwarfs
mostly have face-on central surface brightness $\mu_{0,c}(B) \lesssim
22\, {\rm mag\, arcsec^{-2}}$, while dIs have $\mu_{0,c}(B) \gtrsim 22\,
{\rm mag\, arcsec^{-2}}$.  On average the difference in $\mu_{0,c}(B)$
between dI and starburst dwarfs amounts to $\sim 2.5\, {\rm mag\,
arcsec^{-2}}$.  While measurement bias is a concern (it is difficult to
detect a compact low surface brightness envelope in the presence of a
dominant starburst), these studies have explored enough of parameter
space that low surface brightness envelopes to starburst dwarfs should
have been detected if they existed.  Instead only high surface
brightness envelopes have been found.  The large change in the structure of a 
dI required to transform it into a starburst dwarf envelope makes it
unlikely that that such a transformation can occur as a dynamical
response to the formation of the starburst core.  Hence it is unlikely
that typical dIs are the progenitors of starburst dwarfs.  Instead it appears
that episodes of star formation occur in both low surface brightness and
high surface brightness hosts with the intensity of the star formation
episode correlated with the host surface brightness.  Only in
high surface brightness hosts do these episodes reach a high enough
intensity to be classified as a ``starburst''.  

We believe that the most likely progenitors (and immediate evolutionary
descendants) of the starburst dwarfs are the ``coreless'' galaxies
discussed above (which may be the same population as the members of the
dI class with the highest surface brightness).  These have envelopes
like those of the starburst dwarfs but do not exhibit a very blue
starbursting core. Even though they populate lists of ``starburst''
galaxies, we believe this is a misnomer: they enter catalogs because
they are moderately blue and are actively forming stars, and have much
higher surface-brightnesses than typical dIs. Since the bursts we have
studied are modest events involving only a few percent of the stellar
mass, high-surface-brightness gas-rich dwarfs could cycle between burst
(strong core) and quiescent (weak/absent core) multiple times with a
fairly large duty cycle (cf. \cite{salzetal95}) until they use up or
expel the remaining gas.

Ultimately the gas supply will be exhausted (and the star formation will
then cease). The envelope properties would then fade on a timescale of
Gyr into the region of the $M_B$--log$(\alpha^{-1})$ plane populated by
dEs.  One possible way to shut off star formation would be if the burst
in the core blew away the ISM. This does not appear to be energetically
feasible in a single burst.  In addition the typical disk \HI\ geometry
of dwarf galaxies makes it more difficult to blow out large a large
fraction of the ISM (cf. \cite{dYH94}). An alternate method of turning
starburst galaxies into gas poor dEs is through multiple bursts of star
formation that both consume and expel the ISM until it is totally
gone. If galaxies undergo bursts for about 1/3 of their lifetime (as
suggested by the data in \cite{salzetal95}), most of the galaxies in
this sample could turn their current HI gas into stars in about
$10^{10}$ years. Thus, this evolutionary path is a slow one (typical
starburst dwarfs in the present cosmological epoch will not start down
the final path to dE status any time soon).  Conversely, since
present-day dEs most certainly did evolve from gas-rich dwarfs that
experienced one or more intense bursts of star-formation in the distant
past, these bursts must have been more powerful than the events occurring
in typical present-day dwarfs.

We thank the anonymous referee and the scientific editor, Greg Bothun,
for their thoughtful and useful comments.  This paper benefited from
data and literature searches with NED, the NASA/IPAC Extragalactic
Database, a facility operated by the Jet Propulsion Laboratory, Caltech,
under contract with the National Aeronautics and Space
Administration. This work was supported with NASA grant NAGW-3138.


\end{document}